\documentclass[12pt]{article}

\usepackage{amsmath,graphicx}
\usepackage{epstopdf} 
\usepackage{multirow}
\usepackage{amsfonts}
\usepackage{amssymb}
\usepackage{amscd}

\def\hybrid{\topmargin -20pt    \oddsidemargin 0pt
        \headheight 0pt \headsep 0pt
        \textwidth 6.25in       
        \textheight 9.5in       
        \marginparwidth .875in
        \parskip 5pt plus 1pt   \jot = 1.5ex}

\hybrid
\usepackage{xy}
\input xy 
\xyoption{all}
\usepackage{amsmath}
\usepackage{hyperref}
\usepackage{graphicx}
\usepackage{color}
\usepackage{mathcomp}
\usepackage{amssymb}
\usepackage{esint}
\usepackage{amsmath, amsthm, amssymb}
\usepackage{amsmath}
\usepackage{setspace}
\numberwithin{equation}{section}
\makeatletter

\newcommand{\sla}{\slash\!\!\!}

\def\Deltam{\Delta_m}

\def\Deltamu{\Delta_e}
\def\Deltaphi{\Delta_d}

\def\DPhi{\delta \Phi}
\def\DLambda{\delta \Lambda}

\def\beq{\begin{equation}}
\def\eeq{\end{equation}}
\def\beqa{\begin{eqnarray}}
\def\eeqa{\end{eqnarray}}

\def\im{{\rm Im \,}}
\def\re{{\rm Re \,}}

\def\Tr{{\rm Tr \,}}

\def\cd{{\cal D}}

\def\cj{{\cal J}}

\def\mg{m_{3/2}}
\def\mg2{m^2_{3/2}}

\def\Dsl{\,\raise.15ex\hbox{/}\mkern-13.5mu D} 
\def\rep#1{\mbox{{\bf #1}}}

\def\Phid{\Phi_D}

\def\Kp{\hat K}

\def\Kpp{\tilde K}

\newcommand{\cN}{\mathcal{N}}
\newcommand{\cD}{\mathcal{D}}
\newcommand{\cF}{\mathcal{F}}
\newcommand{\bbR}{\mathbb{R}}

\newcommand{\be}{\begin{equation}}
\newcommand{\ee}{\end{equation}}

\DeclareMathOperator{\SL}{\mathit{SL}}

\DeclareMathOperator{\Es7}{\mathit{E}_{7(7)}}

\newcommand{\Tsub}{O(6,6) \times \SL(2,\bbR)}

\newcommand{\SLE}{\SL(8,\bbR)}
\DeclareMathOperator{\Ex6}{\mathit{E}_{6(2)}}
\newcommand{\mukai}[2]{\big<{#1},{#2}\big>}

\newcommand{\nn}{\nonumber}

\newcommand{\FI}{F_{\rm i}}
\newcommand{\FE}{F_{\rm e}}
\newcommand{\FD}{F_{\rm d}}
\newcommand{\PE}{\partial_{\rm e}}

\begin{document}

\begin{titlepage}
\rightline{\small IPhT-T12/051}
\vskip 1.4cm

\begin{center}

{\Large \bf  ${\cal N}=2$ vacua in Generalized Geometry
}

\vskip 1.2cm

{\bf Mariana Gra{\~n}a and Francesco Orsi }

\vskip 0.4cm

{\em Institut de Physique Th\'eorique,                   
CEA/ Saclay \\
91191 Gif-sur-Yvette Cedex, France}  \\
\vskip 0.1cm
{\tt mariana.grana@cea.fr, francesco.orsi@cea.fr}

\vskip 0.4cm
\vspace{15mm}
\begin{center} {\bf ABSTRACT } \end{center}
\vspace{2mm}

\end{center}
We find the conditions on compactifications of type IIA to four-dimensional Minkowski space to preserve $\cN=2$ supersymmetry in the language of Exceptional Generalized Geometry (EGG) and Generalized Complex Geometry (GCG).  In EGG, off-shell $\cN =2$ supersymmetry requires the existence of a pair of $SU(2)_R$ singlet and triplet algebraic structures on the exceptional generalized tangent space that encode all the scalars (NS-NS and R-R) in vector and hypermultiplets respectively. We show that on shell $\cN=2$ requires, except for a single component, these structures to be closed under a derivative twisted by the NS-NS and R-R fluxes. We also derive the corresponding GCG-type equations for the two pairs of pure spinors that build up these structures.

\vfill

\end{titlepage}

\tableofcontents{}
\begin{flushright}
\end{flushright}
\newpage
\section{Introduction}
\label{Intro}
The study of four-dimensional configurations with reduced supersymmetries is crucial to connect string theory with phenomenology.  Even if in many physically interesting situations supersymmetry is expected to be broken, the scale of supersymmetry breaking can be much lower than the compactification scale, and studying supersymmetric compactifications is a first step towards understanding the non-supersymmetric setups. In particular, supersymmetry has been shown to constrain the  allowed internal geometries to certain specific classes. When no fluxes are turned on, supersymmetric backgrounds of type II supergravity of the form $M_{1,9}=\mathbb{R}^{1,3} \times M_6$ require the internal manifold $M_6$ to be Calabi-Yau \cite{CHSW}. Such manifolds satisfy an  \textit{algebraic} condition, namely the existence of a global section on the spinor bundle over $TM_6$ (\textit{i.e.} there should be a globally defined nowhere vanishing internal spinor), and a \textit{differential} one, that the spinor is covariantly constant. The algebraic condition is necessary in order to recover a supersymmetric ($\cN=2$) effective \textit{theory} in four dimensions, while the differential one is required in order to have a supersymmetric \textit{vacuum}. 
\newline{}
Turning on fluxes on the internal manifold is phenomenologically and mathematically interesting in many respects. They were primarily motivated in light of their potential to solve the problem of moduli stabilization \cite{GKP}.  
Their presence also leads to warped spaces, which are of interest in the Randall-Sundrum scenario \cite{RS}, giving a stringy origin to the hierarchy of scales \cite{GKP}. From the mathematical point of view, while they leave the algebraic constrain intact, vacua with fluxes are possible on manifolds which have weaker differential properties. Rephrasing these constraints in a similar language as those for fluxless solutions was very much guided by the framework of generalized complex geometry developed by Hitchin \cite{Hitchin}. 
\newline{}
Generalized complex geometry was used in  \cite{GMPT1,GMPT2} to  characterize $\cN=1$ vacua. In analogy with the fluxless case, off-shell supersymmetry requires an algebraic condition to hold, namely the existence of a pair of pure spinors on the spinor bundle over the generalized tangent bundle $TM_6\oplus T^*M_6$. These pure spinors geometrize the entire NS-NS content of type II string theories, as they determine the metric, B-field and dilaton.  To describe a vacuum, the pair of pure spinors should also satisfy specific differential conditions \cite{GMPT2}, namely the pure spinor that has the same parity as the R-R fluxes should be closed (and thus the manifold is said to be generalized Calabi-Yau), while the non closure of the second pure spinor is due to the R-R fluxes. Alternatively \cite{BC,KM}, these conditions can be obtained from the F and D-terms of the effective four-dimensional gauged supergravity \cite{GLW,BG}. It has been also proven that the pure spinor equations can be deduced from a generalized calibration condition for D-branes \cite{MS,Mart}.
\newline{}
The R-R fields are not geometrized in the language of generalized complex geometry. Including them in some geometric structure necessarily demands enlarging the generalized tangent bundle, so that it includes the extra charges carried by D-branes. 
The natural generalization appears to be Exceptional (or Extended) Generalized Geometry (EGG) \cite{Hull, PW, W2}, its name alluding to the covariance under the exceptional groups appearing in U-duality. 
\newline{}
The algebraic conditions to have $\cN=2$ supersymmetry in four-dimensions have been worked out in the language of EGG in \cite{GLSW}.  Very much in analogy to the generalized complex geometric case, they require the existence of two algebraic structures on the exceptional generalized tangent bundle (in fact one of them, rather than a single structure, is actually a triplet satisfying an $SU(2)_R$ algebra), which are built by tensoring the internal $SU(8)$ spinors. The $SU(2)_R$-singlet structure, that we call $L$, describes the vector multiplet moduli space, while the triplet of structures (named $K_a$) describes the hypermultiplets.
In type IIA (IIB) the structure $L$ contains a difference of two even (odd) $O(6,6)$ pure spinors, plus extra vectorial degrees of freedom, while the structures $K_a$ contain their odd (even) chirality counterparts, plus an additional bivector, two-form and a couple of scalars.
\newline{}
Differential conditions in order to have an $\cN=1$ vacuum in this language have been studied in \cite{GO1}\footnote{Steps in this direction were done in \cite{GLSW} (see also in \cite{PW} for the M-theory case), where a set of natural $E_{7(7)}$-covariant equations was conjectured to describe $\cN=1$ vacua.}, where it was found that  $\cN=1$ supersymmetry requires on one hand closure of both $L$ and $r^a K_a$, where $r^a$ is a vector pointing in the direction of the $\cN=1$ supersymmetry preserved. On the other hand, the structure along the complex orthogonal direction\footnote{The splitting into parallel and orthogonal directions with respect to $U(1)_R$ is the same as the one used to identify respectively the D-term and superpotential out of the triplet of Killing prepotentials in $\cN=2$ theories.} is closed upon projecting onto the holomorphic sub-bundle defined by $L$. 
\newline{}
The aim of this paper is to investigate the differential conditions on these structures required by $\cN=2$ supersymmetric vacua on four-dimensional Minkowski space, and their corresponding expression in terms of the $O(6,6)$ pure spinors that they contain. 
A generic $\cN=2$ theory possess an $SU(2)_R$ R-symmetry, which must left be unbroken in the $\cN=2$ compactification, and therefore the conditions should be the same for the three $K_a$. We expect that  $\cN=2$ supersymmetry should be translated into integrability of the structures $L, K_a$. We show in this paper that all but one component of the derivative of $L$ and $K_a$ are required to vanish. The two components (one in the derivative of $L$ and one in the derivative of $K$) that do not vanish, involve representations that should be projected out in
order to obtain a standard $\cN=2$ effective four-dimensional supergravity description (\textit{i.e.} a description without massive gravitini multiplets), but are there in the ten-dimensional formulation. We  work in type IIA, though we expect the same equations to hold in type IIB \footnote{The R-R 4-form  flux appears explicitly on the RHS of one of the EGG equations, and it should get appropiately modified in the type IIB case.}. We also write these equations in the language of GCG, \textit{i.e.} we find the equations governing the two pairs of pure spinors that build up $L$ and $K$. These equations involve the twisted differential $d-H \wedge$, and the R-R fluxes appear on the right hand side only when we consider the extra degrees of freedom.  
\newline{}
Conditions for unbroken ${\cal N}=2$ supersymmetry for type IIB compactifications on conformal Calabi-Yau manifolds were obtained in \cite{KST}, by further restricting the $\cN=1$ requirements found in \cite{GP}\footnote{The $\cN=1$ conditions require the 3-form flux $G_3$ to be of type (2,1) and primitive \textit{i.e.} in the $6$ of Calabi-Yau $SU(3)_H$ holonomy. In $\cN=2$, the $SU(2)_R$ symmetry, which embeds in $SO(6)$ as $SU(2)_L\times SU(2)_R\times U(1) \subset SO(6)$, splits the $ 6$  into $3+2+1$ under $SU(2)_L\subset SU(3)_H$. In order to preserve $\cN=2$ supersymmetries, the flux $G_3$ should satisfy a further constraint: it must be in the $3$ representation, or more precisely in the $(3,0)_2$ of $SU(2)_L\times SU(2)_R\times U(1)$.}. On more general manifolds and using the generalized geometric language, our current understanding of the conditions for $\cN=2$ vacua amounts to checking whether there are two pairs of pure spinors giving the same metric, B-field and dilaton, which separately satisfy the $\cN=1$ conditions\footnote{Note that for such thing to happen, the manifold needs to have at least two never parallel globally defined internal spinors, or in other words have $SU(2)$ (or smaller) structure.}. This is how $\cN=2$ solutions have been obtained in \cite{TT} (for their description in terms of $\cN=2$ gauged supergravity see \cite{Andria})  and  \cite{LT2}.  On the other hand, a detailed analysis of the supersymmetric conditions leading to $\cN=2$ AdS$_4$ or Minkowski vacua from a gauged supergravity point of view is done in \cite{HLV,Louis}, which provide concrete examples, some of which in the context of  flux compactifications of M-theory. We will make contact with these works in the discussion. 
\newline{}
The study of $\cN=2$ vacua is interesting also for the applications of the AdS/CFT correspondence in settings with reduced number of supersymmetries. This has for instance been investigated in the circle reduction of $M^{1,1,1}$ giving a massive deformation of the AdS$_4\times M_6$ backgrounds \cite{PZ}, as well as for a first order perturbative expansion in the Romans mass \cite{GT} .
\newline{}
The paper is organized as follows: in Section \ref{GCG} we introduce the necessary concepts of generalized complex geometry. In Section \ref{EGG} we show the main features of the extended or exceptional version of generalized geometry. In Section \ref{diffprev} we present the differential conditions on the algebraic structures required by $\cN=2$ supersymmetry on-shell. In Section \ref{sec:N=2GCG} we write the equations for vacua in terms of pure spinors, and we finish by a discussion in Section \ref{discussion}. Appendix \ref{app:N=1} reviews the generalized complex and exceptional geometric formulation of $\cN=1$ vacua. Appendix \ref{SL8LK} shows the different components of the algebraic structures in terms of pure spinors. Appendix \ref{E7} contains the tensor product formulae needed in computing the derivatives of the algebraic structures. Appendix \ref{susy} gives the equations on the $SU(8)$ spinors obtained from the ten-dimensional supersymmetry transformations, and Appendix \ref{app:DLDKvssusy} includes the details of the derivation of the Eqs. presented in Sections \ref{diffprev} and \ref{sec:N=2GCG}. 

\section{Generalized  complex geometry} \label{GCG}

In Generalized (Complex) Geometry\cite{Hitchin}, one constructs algebraic structures on the generalized tangent bundle $TM\oplus T^*M$.
These structures appear in compactifications of type II theories as they are constructed from the tensor product of two internal spinors. We will concentrate on compactifications of type IIA to four-dimensional warped Minkowski space, \textit{i.e.} the  ten-dimensional metric is
\beq \label{warped}
ds^2=e^{2A} \eta_{\mu\nu} dx^\mu dx^\nu + ds_6^2 \ .
\eeq
In order to recover an ${\cal N}=2$ effective action in four-dimensions, the following splitting of the ten-dimensional spinors should be globally well-defined
\beq \label{specialansatz}
\begin{aligned}
\varepsilon^1&= \zeta^1_- \otimes \eta^1_+ + {\rm h.c.} \\
\varepsilon^2&= \zeta^2_- \otimes \eta^2_{\mp} + {\rm h.c.} 
\end{aligned}
\eeq
where the minus (plus) sign on the chirality of $\eta^2$ is for type IIA (IIB). We will later see that this is not the most general ansatz for the four-six splitting, but in terms of the effective 4D theory, as well as to study ${\cal N}=1$ vacua, one can always make a redefinition such that the splitting has this form. This is not true though when we study ${\cal N}=2$ vacua. We will come back to this point several times in the text.
\newline{}
Tensoring the two internal spinors, one can build Weyl pure spinors\footnote{A spinor is said to be pure if its annihilator space, defined as $L_{\Phi}=\{x+\xi \in TM\oplus T^*M \big| (x+\xi)\cdot\Phi=0 \}$
is maximal (here $\cdot$ refers to the Clifford action $X \cdot \Phi=X_A \Gamma^A \Phi$, A=1,...,12), \textit{i.e.} 6-dimensional in our case.} of $O(6,6)$, namely 
\begin{equation}
\Phi^+=e^{-\phi} \eta^{1}_+ \eta^{2\dagger}_{+}, \qquad \Phi^-=e^{-\phi} \eta^{1}_+ \eta^{2\dagger}_{-}
\label{purespinors}
\end{equation}
where the plus and minus refer to spinor chirality, and $\phi$ is the dilaton, which defines
the isomorphism between the spinor bundle and the bundle of forms. Using 
Fierz identities, these can be expanded as
\begin{equation}
\eta^1_{\pm} \eta^{2\dagger}_{\pm}=\frac{1}{8}\overset{6}{\underset{k=0}\sum}\frac{1}{k!}(\eta^{2\dagger}_{\pm}\gamma_{m_k\dots i_1}\eta^1_{\pm})\gamma^{i_1\dots m_k} \ .
\label{bilinears}
\end{equation}
Using the isomorphism between the spinor bundle and the bundle of differential forms (often referred to as Clifford map):
\begin{equation} \label{Cliffordmap}
 A_{m_1\dots m_k} \gamma^{m_1\dots m_k}  \longleftrightarrow A_{m_1 \dots m_k} dx^{m_1}\wedge\dots\wedge dx^{m_k}
\end{equation}
the spinor bilinears (\ref{bilinears}) can be mapped to sums of forms. Under this isomorphism, the inner product of spinors  
$\Phi \chi$ is mapped to the following action on forms, called the Mukai pairing
\begin{equation} \label{Mukai}
\langle\Phi,\chi\rangle=(\Phi\wedge s(\chi))_6, \quad  \,\,\mbox{where     } s(\chi)=(-)^{\mbox{\begin{scriptsize}Int\end{scriptsize}}[n/2]}\chi
\end{equation}
and the subindex 6 means the six-form part of the wedge product.\\
For Weyl $O(6,6)$ spinors, the corresponding forms are only even (odd) for
a positive (negative) chirality $O(6,6)$ spinor. 
In the special case where $\eta^1=\eta^2\equiv\eta$, familiar from the case of Calabi-Yau compactifications, we get
\beq \label{Phipmsu3}
\Phi^+=e^{-\phi} e^{-iJ} \ , \qquad \Phi^-=-i e^{-\phi} \Omega
\eeq
where $J, \Omega$ are respectively the symplectic and complex structures of the manifold.  
Pure spinors can be ``rotated" by means of $O(6,6)$ transformations. Of particular interest is the nilpotent
subgroup of $O(6,6)$ defined by the generator
\beq \label{Btransform}
{\cal B}=  \begin{pmatrix}
            0 & 0 \\  B & 0        
            \end{pmatrix} \ , 
\eeq 
with $B$  an antisymmetric $6\times6$ matrix, or equivalently a two-form. On spinors it amounts to the exponential action 
\beq \label{Btwistedspinors}
\Phi^\pm \to e^{-B} \Phi^\pm \equiv \Phid^\pm  
\eeq
We will refer to $\Phi$ as naked pure spinor, while $\Phid$ will be called dressed pure spinor.
The pair ($\Phid^+,\Phid^-$) defines a positive definite metric on the generalized tangent space,
which in turn defines a positive metric and a two-form (the $B$ field) on the six-dimensional manifold. \\
In an analogous way as an $O(6)$ spinor defines an $SU(3)$ structure (\textit{i.e.}, it is invariant under an $SU(3)$ subgroup of
$O(6)$), a pure $O(6,6)$ spinor defines an $SU(3,3)\subset O(6,6)$ structure. Its 32 degrees of freedom minus one corresponding to the norm parameterize the coset $O(6,6)/SU(3,3)$. Furthermore, two $O(6)$ spinors
which are never parallel, define an $SU(2)$ structure, which is the intersection of the two $SU(3)$ structures. Similarly, two pure $O(6,6)$ spinors, whenever they satisfy the following compatibility condition
\begin{equation}
\langle \Phi^+,\Gamma^A\Phi^-\rangle=0, \quad A=1,\dots,12 \ ,
\end{equation} 
define an $SU(3) \times SU(3)$ structure. Pure spinors which are tensor products of $O(6)$ spinors as defined in (\ref{purespinors}) are automatically compatible. 
\newline{}
We finish this section by mentioning that the 6d annihilator space of an $O(6,6)$ pure spinor can be thought as the holomorphic bundle of a generalized almost complex structure  (GACS) $\cj$, which  is a map
from $TM\oplus T^*M$ to itself such that it satisfies the hermiticity condition ($\mathcal{J}^t\eta\mathcal{J}=\eta$) and $\mathcal{J}^2=-1_{12}$. Therefore there is a one-to-one correspondence between a pure spinor of $O(6,6)$ and a GACS.  The GACS can be obtained from the pure spinor by \cite{GLW}\footnote{The correspondence is actually many-to-one since rescaling the pure spinor by a complex number gives rise to the same GACS\label{foot:mto}.} 
\begin{equation}
\label{Jgen}
   \mathcal{J}^{\pm A}{}_B\ =\ i \, \frac{
       \mukai{\bar \Phi^\pm}{\Gamma^A{}_B {\Phi}^\pm}}
       {\mukai{\Phi^\pm}{\bar{\Phi}^\pm}}\ .
\end{equation}

\section{Exceptional Generalized Geometry} \label{EGG}
To incorporate the R-R fields to the geometry, in exceptional generalized geometry (EGG) \cite{Hull,PW} one extends the tangent space (or rather the generalized tangent space $T \oplus T^*$) such that  there is a natural action of  the U-duality group on it. In this paper we will be interested in compactifications of type II theories (and in particular we will work with type IIA\footnote{Many things can be easily translated to type IIB by switching chiralities.}) on six-dimensional manifolds, where the relevant exceptional group is   $E_{7(7)}$. 
Shifts of the B-field a well as shifts of the sum of internal R-R fields $C^{-}=C_{1}+C_{3}+C_{5}$ correspond to particular $E_{7(7)}$ adjoint actions. To form a set of gauge fields that is closed under U-duality, we also have to consider the shift of the six-form dual of $B_2$, which we will call $\tilde B$. \footnote{Equivalently these are shifts of the dual axion $B_{\mu\nu}$.}  \\
In what follows, we will mainly use the decomposition of $E_{7(7)}$ under $\SLE$. This subgroup contains the product $SL(2,\mathbb{R})\times GL(6,\mathbb{R})$, and allows to make contact with $SU(8)/ \mathbb{Z}_2$, the maximal compact subgroup of $E_{7(7)}$. The latter is the group under which the spinors transform, and therefore the natural language to formulate supersymmetry using the Killing spinor equations. 
\newline{}
In our analysis we will use the fundamental ${\bf 56}$, the adjoint ${\bf 133}$ and the ${\bf 912}$ representations of $\Es7$. The first one decomposes under $\SLE$ as
\begin{align} 
\bf{56}&=\bf{28}+{\bf{28'}} \label{slf} \quad  \\
\nu&=(\nu_{ab}, {\nu'}^{ab})  \ 
\end{align}
where $a,b=1,...,8$ and $\nu_{ab}=-\nu_{ba}$. We will also denote $6d$ coordinates by $m,n=1,...,6$ and $SL(2,\mathbb{R})$ indices by $i,j=1,2$.
\newline{}
The adjoint has the following decomposition  
\begin{align}
\bf{133}&=\bf{63}+\bf{70}\label{sla} \\
\mu&=(\mu^{a}_{\,\,\,b},\mu_{abcd}) \nonumber
\end{align}
where the first piece corresponds to the adjoint of $\SLE$, and we have $\mu^a{}_a=0$ and $\mu_{abcd}$ is fully antisymmetric. We also define $\mu^{abcd}\equiv\frac{1}{4!} \epsilon^{abcdefgh} \mu_{abcd}=(\star \mu)^{abcd}$ \footnote{\label{foot:Hodge}We use $\star$ for the eight-dimensional Hodge dual, while $*$ refers to the six-dimensional one.}.
\newline{}
For the \textbf{912} we have
\begin{align}
\bf{912}&=\bf{36}+\bf{420}+\bf{36}'+\bf{420}'\label{ntsl8} \\
\phi&=(\phi_{ab},\phi_{abc}{}^{d},\phi'^{ab},\phi'^{abc}{}_{d}) \nn 
\end{align}
with $\phi_{ab}=\phi_{ba}$, $\phi_{abc}{}^d=\phi_{[abc]}{}^d$ and $\phi_{abc}{}^c=0$. 
\newline{}
The fundamental representation is where the charges live. Momentum and winding charges are embedded respectively in $\nu'^{2m}$ and $\nu_{1m}$, while D0, D2, D4 and D6-brane charges in $\nu'^{12}$, $\nu_{mn}$, $\nu'^{mn}$ and $\nu_{12}$ (for more details, see \cite{GO1}). 
\newline{}
The gauge fields live in the adjoint representation. Their embedding in terms of the $SL(8,\mathbb{R})$ components in (\ref{sla}) is the following \cite{GO1}  
\begin{align} \label{calAsl8}
\mu^1{}_2&=\tilde B \ , \qquad \mu^1{}_m=-C_m \ , \qquad \mu^m{}_2=(*C_{(5}))^m \ ,\\
 \mu_{mn12}& =-B_{mn} \ , \qquad \mu_{mnp2}=-C_{mnp} \ \nn . 
\end{align}
 \newline{}
Finally, the fluxes live in the ${\bf 912}$, and are embedded as\footnote{The factors of the dilaton in these formulae appear because  on one hand we are using an eight-dimensional metric of the form written in (\ref{8dmetric}), and on the other we have to consider the ${\bf 912}$ representation weighted by the factor $g^{-1/4}e^{-\phi}$. Then $\phi'^{11}$, for instance, transforms as $e^{\phi} \otimes {\mathbb R}$. For more details see Appendix \ref{SL8LK}.}
\begin{align} \label{sl8F}
\phi_{22}&=e^{\phi}(*F_{6}) \ , \qquad
 \phi_{mn2}^{\,\,\,\quad1}=-\frac{e^{\phi}}{2}F_{mn}\nn \\
\phi'^{11}&= e^{\phi}F_0 \ , \qquad \phi'^{mnp}{}_{2}=-\frac{1}{2}(*H)^{mnp} \ , \qquad
\phi'^{mn1}{}_{2}=-\frac{e^{\phi}}{2} (*F_4)^{mn} \ .
\end{align}

\subsection{$E_{7(7)}$ structures as spinor bilinears} 
The supersymmetry parameters transform under the maximal compact subgroup of the duality group, which in the case at hand is  $SU(8)$. The action of this group on the spinors\footnote{For conventions on spinors see Appendix B of \cite{GO1}.} is manifest if we combine the two ten-dimensional supersymmetry $\epsilon^1, \epsilon^2$ as follows
\begin{align}
\left(\begin{array}{c}\epsilon^1\\ \epsilon^2\end{array}\right)=\zeta^1_-\otimes \theta^1 +\zeta_-^2\otimes \theta^2 +\mbox{h.c.}\label{suvar}
\end{align}
where $\zeta^{1,2}_-$ are four-dimensional spinors of negative chirality, and $\theta^{1,2}$ are never parallel, and can be parameterized as
\begin{equation}
\theta^1=\left(\begin{array}{c} 
\eta^{1}_{+}\\
\tilde{\eta}^1_{-}
\end{array}\right), \qquad
\theta^2=\left(\begin{array}{c} \tilde{\eta}^{2}_{+}\\
\eta^2_{-} \ 
\end{array}\right)\label{fulla} \ .
\end{equation}
The notation is chosen so that the standard ansatz for the ten-dimensional spinors (\ref{specialansatz}) (used for example in \cite{GLW}) corresponds to $\tilde \eta^1=\tilde \eta^2=0$. Introducing $\tilde \eta^I$ gives us the most general ansatz for four-dimensional ${\cal N}=2$ supersymmetry. 
\newline{}
A nowhere vanishing spinor $\theta$ defines an $SU(7)\subset SU(8)$ structure. The pair $(\theta^1,\theta^2)$ defines an $SU(6)$ structure\footnote{Note that an $SU(6)$ structure can be built out of  a single globally defined internal spinor $\eta$, taking  $\eta^1=\eta^2=\eta$, $\tilde \eta=0$. However, this type of $SU(6)$ structure (for which many of the components of $L$ and $K$ defined in (\ref{lambdaK}) vanish), does not admit ${\cal N}=2$ vacua with non-vanishing fluxes.}. Without loss of generality, we can choose a basis where the spinors are orthonormal, namely  
\beq \label{normtheta}
\bar \theta_I \, \theta^J= \delta_I{}^J \ .
\eeq 
where $I=1,2$ is a fundamental $SU(2)_R$.
 \newline{}
The two spinors can be combined into the following $SU(2)_R$ singlet and triplet structures, which replace the pure spinors of GCG, and parameterize respectively the scalars from vector multiplets and hypermultiplets
\beq \label{lambdaK}
L=e^{-\phi} \epsilon_{IJ}\theta^{I}\theta^{J} \ , \qquad K_{a} =\frac{1}{2} e^{-\phi} \sigma_{aI}{}^J\theta^{I}\bar{\theta}_{J} \ ,  
\eeq
 The triplet
$K_a$ satisfies the $su(2)$ algebra with a scaling given by the dilaton, \textit{i.e.}
\beq
[K_a , K_b]= 2  e^{-\phi} \epsilon_{abc} K_c \ .
\eeq
$L$ and $K_a$ are the $E_{7(7)}$ structures that play the role of the generalized almost complex structures $\Phi^+$ and $\Phi^-$.
They belong respectively to the $\rep{28}$ and $\rep{63}$ representations of $SU(8)$, which are in turn part of the $\rep{56}$ and $\rep{133}$ representations of $E_{7(7)}$. 
\newline{}
We will use the $SL(8,\mathbb{R})$ decomposition of $L$ and $K_a$. The former is obtained from the $SU(8)$ object in (\ref{lambdaK}) by
\beq\label{pw1}
L^{ab}=\lambda'^{ab}+i  \lambda^{ab}=\frac{\sqrt 2}{4} L^{\alpha\beta} \Gamma^{ab}{}_{\beta\alpha}
\eeq
where $\lambda$ and $\lambda'$ are respectively the ${\bf 28}$ and ${\bf 28}'$ (real) components of $L$, and $\alpha, \beta=1,...,8$ are Spin(8) spinor indices. As for $K_a$, given that it is in the ${\bf 63}$ representation of $SU(8)$, we get that its $SL(8,\mathbb{R})$ components are
\begin{align}
K^{ab}&=-\frac{1}{4} K^{\alpha}{}_{\beta} \Gamma^{ab}{}^{\beta}{}_{\alpha} \nonumber\\
K_{abcd}&= \frac{i}{8} K^{\alpha}{}_{\beta} \Gamma_{abcd}{}^{\beta}{}_{\alpha}  \label{pw2}
\end{align}
 where $K^{ba}=-K^{ab}$ (and $K^{ab}=K^a{}_c \, \hat g^{cb}$) and $\star K_{abcd}=-K_{abcd}$
(the symmetric and self-dual pieces would be obtained from the $\rep{70}$ representation $K^{\alpha\beta\gamma\delta}$, which is not there).
\newline{}
We give in Appendix \ref{SL8LK} the different $SL(8,\mathbb R)$ components of $L$ and $K_a$ in terms of bilinears of the 6d spinors $\eta^I, \tilde \eta^I$ in (\ref{fulla}) that build up $\theta^I$.
\newline{}
The structures $L$ and $K_{a}$ can be dressed by the action of the gauge fields $B$, $\tilde B$ and $C^-$,  \textit{i.e.} we define
\beq
\label{fulllambdaK}
L_D=e^{C} e^{\tilde B} e^{-B}  L \ , \qquad K_{aD}=e^{C} e^{\tilde B} e^{-B}  K_{a} \ ,
\eeq
where the action of $C, \tilde B, B$ on $L$ and $K$ is given respectively by (\ref{56x133=56SLE}) and (\ref{133x133=133SLE}) and we have to use their embedding in $\Es7$, given in (\ref{calAsl8}). 
They span orbits in $E_{7(7)}$ which are respectively Special K\"ahler and
quaternionic. As shown in \cite{GLSW}, the structure $L_D$ is stabilized by $\Ex6$, and the corresponding local 
Special K\"ahler space is $\frac{E_{7(7)}}{\Ex6} \times U(1)$. The triplet $K_{aD}$ is stabilized by an $SO^*(12)$
subgroup of $E_{7(7)}$, and the corresponding orbit is the quaternionic space $\frac{E_{7(7)}}{SO^*(12) \times SU(2)}$, where
the $SU(2)$ factor corresponds to rotations of the triplet. The $SO^*(12)$ and $\Ex6$ structures intersect on an
$SU(6)$ structure if $L$ and $K_a$ satisfy the compatibility condition
\beq
L \cdot K_a\big|_{\mathbf{56}}=0 \ ,
\eeq
where we have to apply the projection on the $\rep{56}$ on the product $\rep{56} \times \rep{133}$. This condition is automatically satisfied for the structures (\ref{lambdaK}) built up as spinor
bilinears.
\section{Conditions for ${\cal N}=2$ flux vacua}\label{diffprev}
In this section we determine the equations for the structures \eqref{lambdaK} required by  ${\cal N}=2$ supersymmetric compactifications on warped Minkowski space, \textit{i.e.} where the ten-dimensional metric has the form (\ref{warped}).
From the point of view of the effective four-dimensional ${\cal N}=2$ action, these come from setting to zero the triplet of Killing prepotentials $P_a$, along with its variations. In the language of EGG, the triplet of Killing prepotentials reads \cite{GLSW} 
\beq \label{PaE7}
P_a=S(L,{\cal D} K_a)
\eeq
where $S$ is the symplectic invariant on the ${\bf 56}$ representation, given in (\ref{symplSLE}) and  ${\cal D}$ is the  derivative twisted by the fluxes constructed as explained below. From demanding that this is zero under variations of $L$ and $K_a$, one expects that ${\cN}=2$ supersymmetry requires that both $L$ and the whole triplet $K_a$ are closed under ${\cal D}$.
 We will see  that this is roughly the case, though some subtleties arise. But before presenting the equations on $L$ and $K_a$, we will explain very briefly, following \cite{GO1}, how the twisted derivative is built and how it acts.  
\newline{}
We define 
\beq \label{calD}
{\cal D}=D+ {\cal F}
\eeq 
where the derivative $D$ is in the $\rep{56}$ representation, and its $\SLE$ decomposition (see (\ref{slf})) is given by
\beq \label{LCsl8}
D_{m2}= \nabla_m   \ ,
\eeq
(all otheer components are zero), and the fluxes ${\cal F}$ are in the ${\bf 912}$ representation, and are given in terms of $\SLE$ in (\ref{sl8F}) \footnote{The fluxes are obtained by ${\cal F}=e^B e^{-\tilde B} e^{-C} D \, e^{C} e^{\tilde B} e^{-B}\big|_{\bf{912}}$.}. 
\newline{}
The equations involve the twisted derivatives of $L$ and $K$ projected onto specific representations, respectively the ${\bf 133}$ and ${\bf 56}$. We therefore have to use the following tensor products 
\begin{align}
{\cal D} L= (\quad \quad D L\ \, \ \ \ &+ \ \ \  {\cal F} \, L \, \quad )|_{\bf 133}, \label{DLtensorprod}  \\
\rep{56} \times \rep{56}|_{\bf 133} \ &+ \ \rep{912} \times \rep{56}|_{\bf 133}\nn
\end{align}
\begin{align}
{\cal D} K= (\quad  \quad D K\ \, \ \ \ &+ \ \ \  {\cal F} \, K \quad)|_{\bf 56} \label{DKtensorprod} \\
 \qquad \qquad \qquad\rep{56} \times \rep{133}|_{\bf 56}  \,\,&+ \ \rep{912} \times \rep{133}|_{\bf 56} \nn
\end{align}
 All the formulae for these tensor products are given in Appendix \ref{E7}.
 \newline{}
We will show that ${\cal N}=2$ supersymmetry requires on $L$
\begin{align} \label{DL}
 (\mathcal{D}(e^{\phi}L))^1{}_1 &=0\ ,   \nn \\
 (\mathcal{D}(e^{\phi}L))^2{}_2 &=0\ ,   \nn \\
(\mathcal{D}(e^{-\phi}L))^1{}_2&=0 \ , \nn \\
(\mathcal{D} L)^{1}{}_m&=0\, , \\
(\mathcal{D} L)^{m}{}_2&=0\, , \nn  \\
(\mathcal{D} L)_{mnp2}&=0\, , \nn \\
 (\mathcal{D}(e^{\phi}L))_{mn12}&=0\,, \nn \\
 (\mathcal{D}(e^{\phi-A}L))^{n}{}_{m}&= - \frac{i}{4} e^{2\phi-A} F^{n}{}_{mpq} L^{pq} \ , \nn 
\end{align}
while the other components are trivially zero.
\newline{}
On $K_a$, supersymmetry requires the following equations on any of them\footnote{As expected, the equations are invariant under $SU(2)_R$.} 
\begin{align} \label{DK}
(\mathcal{D} \Kp_a)_{mn}&=0\,\nn \\
({\mathcal{D} \Kp_a})'_{mn}&=0\,\nn \\
(\mathcal{D} \Kp_a)_{12}&=0\,\nn \\
({\mathcal{D} \Kp_a})'_{12}&=0\,\nn \\
(\mathcal{D} (e^{-\phi}\Kp_a))'_{m1}&=0\,,\\
(\mathcal{D} ({e^{\phi}\Kp_a}))_{m1}&=0\,,\nn \\
(\mathcal{D} ({e^{-(2A+\phi)}\Kp_a}))_{m2}&=-e^{-(2A+\phi)}H_{mpq}\Kp_a^{12pq}\,.\nn
\end{align}
where the remaining component ($({\mathcal{D} \Kp_a})'_{m2}$) is trivially zero, and we have defined
\be
\Kp_a \equiv e^{A}K_a\label{kn2}
\ee
and the prime indicates the ${\bf 28'}$ representation of $\SLE$ (see decomposition in (\ref{slf})), whose indices have been lowered with the 8d metric given in \eqref{8dmetric}. The powers of the dilaton in this metric explain the different  powers of the dilaton appearing in these equations, as we will show later.   
\newline{}
We will now briefly show how we obtained these equations, leaving the full details to Appendix \ref{app:DLDKvssusy}. 

\subsection{Explicit form of the twisted derivative}\label{n2}
The twisted derivative defined in (\ref{calD}), applied to $L$ and projected onto the ${\bf 133}$ representation as in (\ref{DLtensorprod}) (where the tensor products needed are given in \eqref{56x56=133SLE} and \eqref{912x56=133SLE} in terms of $\SLE$ decompositions), gives the following components\footnote{Here we are giving the equations for a complex ${\bf 28}$ object as defined in (\ref{pw1}).}  
\begin{align}
(\mathcal{D}L)^{1}_{\,\,\,1}&=-\frac{1}{4}\nabla_{p}L^{p2}\label{Lfirst}\\
(\mathcal{D}L)^{2}_{\,\,\,2}&=\frac{3}{4}\nabla_{m}L^{m2} \label{DL22} \\
(\mathcal{D}L)^{1}_{\,\,\,2}&=-\nabla_{m}L^{1m}-e^{\phi} (*F_{6}) L^{12}-ie^{\phi} F_{0}  L_{12}
+\frac{e^{\phi}}{2}F_{mn}L^{mn}  \nn\\&+i \frac{e^{\phi}}{2}(*F_4)^{np}L_{np}  \label{DL12} \\
(\mathcal{D}L)^{m}{}_{2}&=-\nabla_{p}L^{mp}+\frac{i}{2}(*H)^{mnp}L_{np} -e^{\phi} (*F_{6}) L^{m2}+ie^{\phi} (*F_4)^{mn}   L_{n1} \label{DLm2}\\
(\mathcal{D}L)^{1}_{\,\,\,m}&=\nabla_{m}L^{12}-ie^{\phi} F_{0}  L_{1m}+e^{\phi} F_{mn} L^{n2}\label{lt1m} \\
(\mathcal{D}L)^{n}_{\,\,\,m}&= \nabla_{m}L^{n2} -\frac{1}{4} g^n{}_m \nabla_{p}L^{p2} \label{DLnm}\\
(\mathcal{D}L)_{mnp2}&= \frac{3i}{2}\nabla_{[m} L_{np]}+\frac{1}{2}H_{mnp}L^{12}+\frac{3}{2}ie^{\phi} F_{[mn|}  L_{|p]1}-\frac{e^{\phi}}{2}F_{mnpq} L^{2q} \label{DLmnp2}\\
(\mathcal{D}L)_{mn12}&=i\nabla_{[m} L_{n]1}+\frac{1}{2}H_{mnp}L^{p2}  \ .\label{Llast}
\end{align}
\newline{}
On the other hand, for $K$ we use (\ref{DKtensorprod}), and the tensor products in (\ref{56x133=56SLE}) and (\ref{912x133=56SLE}) and get the following $SL(8,\mathbb{R})$ components 
  \begin{align}
 (\mathcal{D} K)'^{mn}&=-2\nabla_p K^{mnp2}+(*H)^{mnp} K^{2}{}_{p}+e^{\phi} (*F_{4})^{mn} K^{2}{}_{1} \label{Kfirst} \\
(\mathcal{D} K)_{mn}&=-2\nabla_{[m} K^{2}{}_{n]}+e^{\phi} F_{mn}K^{2}{}_{1} \label{DKmn} \\
 (\mathcal{D} K)'^{m1}&=2\nabla_p K^{mp12}+e^{\phi} F_{0}K^{m}{}_{1}- e^{\phi} (*F_{4})^{mn}K^{2}{}_{n}-e^{\phi} F_{np}K^{2npm}  \label{DKpm1}\\
(\mathcal{D} K)_{m1}&=-\nabla_mK^{2}{}_{1} \label{DK5}  \\
 (\mathcal{D} K)'^{m2}&=0  \label{DKpm2}\\
 (\mathcal{D} K)_{m2}&=-\nabla_pK^{p}{}_{m}- H_{mpq} K^{pq12}-e^{\phi} (*F_{6})K^{2}{}_{m}-e^{\phi} F_{mp}K^{p}{}_{1}   \nn \\
 & \quad + e^{\phi} (*F_{4})^{pq}  K_{1pqm}  \label{DKm2} \\
  (\mathcal{D} K)'^{12}&=-e^{\phi} F_{0} K^{2}{}_{1} \label{dkt12u}\\
(\mathcal{D} K)_{12}&=-\nabla_nK^{n}{}_{1}-\frac{1}{3}H_{npq}K^{2npq} -e^{\phi} (*F_{6})K^{2}{}_{1} \label{dkt12l}
 \end{align}

\subsection{Comparing to equations coming from Killing spinors}
As shown in detail in Appendix \ref{susy}, the supersymmetry variations of the internal and external gravitino and the dilatino give algebraic and differential conditions on the ten-dimensional spinors $\varepsilon^1$, $\varepsilon^2$. Using the splitting into four and six-dimensional spinors corresponding to ${\cal N}=2$ supersymmetry, given in (\ref{fulla}), these turn into conditions on the spinors $\theta^I$. We give these conditions in (\ref{intgravitino})-(\ref{dilatino}). Multiplying these by $\theta^J$ (or $\bar \theta_J$) we get conditions on $L$ and $K$ that we give in Appendices \ref{n2L} and \ref{n2K} respectively. 
\newline{}
We show for instance how supersymmetry implies that Eq. \eqref{lt1m} should vanish (this condition is written on the second line in (\ref{DL})): multiplying \eqref{igln2} by $\Gamma^{12}$, as well as $l_d$ times \eqref{dln2} by $i\, \Gamma_m$, and tracing over spinor indices, we recover
\begin{align}
0=&\frac{\sqrt{2}}{4} \Tr \left[ \Gamma^{12} \Delta_m  L+i   \Gamma_{m} l_d \Delta_dL \right] \nonumber\\
=&+\nabla_m L^{12}+\partial_m\phi L^{12}-l_d\partial_m\phi L^{12}\nn\\
&+ \frac{i}{4}H_{mnp}L^{np}(-1+l_d)\nn\\
&+\frac{e^{\phi}}{4}\left[iF_0 (1-5l_d)+(*F_6)(1-l_d)\right]L^1_{\,\,\,m} \nn \\
& +\frac{e^{\phi}}{4}\left[F_{mp}(-1-3l_d)+i(*F_4)_{mp}(1-l_d)\right]L^{2p}\nn\\
=&\nabla_{m}L^{12}-ie^{\phi}F_0 L^{1}{}_{m}-e^{\phi}F_{mp}L^{2p}\nn\\
=&(\cD L)^{1}{}_{m}\,.
\end{align}
where for the third equality we have taken $l_d=1$.
The calculations for the other components of the derivative of $L$ are given in Appendix \ref{n2L}.
\newline{}
We now show how one of the conditions on the derivative of $\Kp$ (defined in \eqref{kn2}), namely the second one in (\ref{DK}), can be recovered by using supersymmetry.
 Taking \eqref{intgravitinoK}  multiplied by $\Gamma^{p1}$ and summing over internal indices, together with $n_e$ times \eqref{egra1} and $n_d$ times \eqref{dila1} multiplied by $-i\Gamma^2$, and tracing the overall sum over the spinor indices, we get
\begin{align} \label{dk12up}
0=&-\frac14 \Tr\left[-\Delta_p \Kp \Gamma^{p1}-i\Gamma^2(n_d\Delta_d+n_e\Delta_e)\Kp\right]\nn\\
&-\nabla_p\Kp^{p1}+\partial_p(A-\phi)\Kp^{p1}-\partial_p(n_e A +n_d \phi)\Kp^{1p}-\frac{1}{2}(1-\frac{n_d}{3})H_{mnp}\Kp^{2mnp}\nn\\
&+\frac{e^{\phi}}{4}\left(iF_0(5n_d+n_e)+(*F_6)(6+n_e-n_d)\right) \Kp^{12} \nn\\
&-\frac{e^{\phi}}{4}\left(F_{mn}(-2+3n_d+n_e)+i\left(*F_4)_{mn}(n_e+n_d\right)\right) \Kp^{mn} 
\end{align}
by choosing here $n_d=1,\,n_e=-1$, we recover
\begin{align}
0=&-\nabla_{p}\Kp^{p1}-\frac{1}{3}H_{mnp}\Kp^{2mnp}+e^{\phi}\big[iF_0+(*F_6)\big]\Kp^{12}\label{proofk2}
\end{align}
In order for the equality to hold, we can further decompose \eqref{proofk2} in terms of its real and imaginary parts, giving respectively\footnote{Actually in our conventions $K^{abcd}, K^{a}{}_b$ are purely imaginary, so the terms real and imaginary should strictly speaking be exchanged.} 
\begin{align}
0=&-\nabla_{p}\Kp^{p1}-\frac{1}{3}H_{mnp}\Kp^{2mnp}+e^{\phi}(*F_6)\Kp^{12}=(\cD \Kp)'^{12}\,,
\end{align}
and
\begin{align}
0\,=&\,e^{\phi}F_0\Kp^{12}=(\cD \Kp)^{12}\,.
\end{align}
For the $mn$ components, a similar argument holds, while for the other equations a slightly more involved calculation is needed.
We show in Appendix \ref{n2K} how to obtain the rest of the conditions for ${\cal D}K$ . 
\newline{}
We note that the equations that have an explicit power of the dilaton in \eqref{DL} are those that involve a derivative of a component of $L$ with one internal and one $SL(2.\mathbb R)$ index. For instance $({\cal D}L)^1{}_1$ is proportional to $\nabla_p L^{p2}$. According to the metric in (\ref{8dmetric}), this component of $L$ transforms as $e^{-\phi} \otimes TM$ (see (\ref{compLSL8})), and this power of the dilaton is compensated by the explicit $e^{\phi}$ factor appearing in the first equation in (\ref{DL}). On ${\cal D} K$, pieces that involve a derivative of a component of $K$ that transforms with a power of  $e^{-\phi}$ (such as $K^{mnp2}$, for example, as shown in \ref{compKSL8})) do not carry explicit dilaton factors, while otheer powers are compensated by explicit powers of the dilaton. For example $({\cal D} K)_{m1}$ contains $\nabla_m K^2{}_1$, which transforms as $e^{-2\phi}$, and this is compensated by the explicit $e^{\phi}$ on the fourth line of (\ref{DK}). 

\section{Conditions for ${\cal N}=2$ vacua in GCG} \label{sec:N=2GCG}
Using the splitting of $\theta^I$ in terms of $SU(4)$ spinors $\eta^I, \tilde \eta^I$ as in (\ref{fulla}), we can obtain $L$ and $K_+=K_1+i K_2$ in terms of $O(6,6)$ pure spinors, namely
\begin{align}
L=\left(\begin{array}{cc} \Lambda^-- \Lambda^{-T} & \Phi^{+}- \bar{\tilde{\Phi}}^{+T}\\
\bar{\tilde{\Phi}}^{+}-{\Phi}^{+T}& \bar \Lambda'^- - \bar \Lambda'^{-T}
\end{array} \right) \ , \qquad
K_{+}= \left(\begin{array}{cc}
\Lambda^+ & \Phi^{-}\\
\bar{\tilde \Phi}^{-} & \bar \Lambda'^+
\end{array}\right) \ ,\end{align}
where here $\tilde \Phi^+$ is defined in an analogous way as $\Phi^+$, Eq. (\ref{purespinors}), but using $\tilde \eta$; the superscript $T$ denotes the transpose of the bispinor and  we have defined
\beq \label{Lambda}
\Lambda^\pm =e^{-\phi} \eta^1_+\tilde \eta^{2 \dagger}_\pm  \ , \quad \Lambda'^\pm =e^{-\phi} \tilde\eta^1_+ \eta^{2 \dagger}_\pm \ .
\eeq
The normalization condition (\ref{normtheta}) implies
\beq \label{normeta}
\eta^{I\, \dagger}_+ \eta^I_++ \tilde\eta^{I\, \dagger}_+ \tilde\eta^I_+=1 \ , \quad \tilde\eta^{2\,\dagger}_+\eta^1_+ + \eta^{2\,\dagger}_-\tilde \eta^1_-=0 \ .
\eeq
Note that the second condition is equivalent to $\Lambda_0^+ + \bar{\Lambda'}_0^{+}=0$, where the subindex $0$ denotes the zero-form component.
\newline{}
The structures $L$ and $K_+$ contain two pure spinors $\Phi$ and $\tilde \Phi$ of positive and negative chirality respectively, plus extra degrees of freedom involving bilinears between  $\eta$ and $\tilde \eta$ (which are zero in the ``standard $\cN=2$ ansatz" introduced in \ref{specialansatz}). In the case of  $K_+$, the two pure spinors $\Phi^-$ and $\tilde \Phi^-$ appear as independent degrees of freedom (unlike $\Phi^+$ and $\tilde \Phi^+$ in $L$). 
\newline{}
In order to get the $SL(8,{\mathbb R})$ components of $L$ and $K$ we use (\ref{pw1}), (\ref{pw2}) and the decomposition of the Gamma matrices in (\ref{gammabasis}). The result is given in (\ref{compLSL8}) and (\ref{compKSL8}). We can see clearly that the extra degrees of freedom in $L$ are ``vectorial" type (\textit{i.e.}, in ${\bf 6}$ representations of $O(6)$, or in terms of the $\Tsub$ subgroup of $\Es7$ they are in the $({\bf 2}, {\bf 12})$), while the extra degrees of freedom in $K$ are in the adjoint of $SL(2)$ and the adjoint of $O(6,6)$.
\newline{}
It is useful to define the polyforms 
\beq \label{DPhi}
\DPhi^+\equiv \sum_{n=0}^{3} \Phi^+_{2n} - (-1)^{[n/2]} \bar{\Tilde \Phi}^+_{2n} \ . 
\qquad \DPhi^- \equiv  \sum_{n=0}^{3} \Phi_{2n+1}^-+ (-1)^{n} \bar{\Tilde \Phi}^-_{2n+1}
\eeq
Using the explicit form of the twisted derivatives in (\ref{Lfirst})-(\ref{dkt12l})  we get that conditions (\ref{DL}) on $L$ imply the following equation on $\DPhi^+$ \footnote{The one, three and five-form pieces come respectively from $({\cal D} L)^1{}_m,({\cal D} L)_{mnp2}$ and $({\cal D} L)^m{}_2$, } 
\beq
\label{dDPhiplus}
d_H \, \DPhi^+ =- 2 \,  \Lambda_1 \cdot F \ . \\
\eeq
where $d_H=d-H\wedge$ and we have defined the polyform and the Clifford action
\beq
\Lambda_1 \cdot = \re \Lambda_1 \llcorner + i \, \im \Lambda_1 \wedge
\eeq
\textit{i.e.} in the $n+1$-form equation in (\ref{dDPhiplus}), the real part of $\Lambda_1$ acts as a vector contracted on $F_{n+2}$, while the imaginary part is a one-form wedged on $F_n$.   
\newline{}
From equations (\ref{DK}) 
specialized to $K_+$, we get \footnote{The two, four and six-form pieces on the second equation come from $({\cal D} K)_{mn},({\cal D} K)'^{mn}$ and $({\cal D} K)_{12}$, and we have used the normalization condition (\ref{normeta}) to express the RHS in terms of $\Lambda_0$.}
\beq 
\label{dDPhiminus}
d_H (e^{A-\phi} \DPhi^-) = -2e^{A-\phi} \, \Lambda_0 \, F  \ .
\eeq
Before writing the additional equations on the other degrees of freedom that appear in $L$ and $K$, we note that these equations involve sums and differences between $\Phi$ and $\tilde \Phi$. While this is expected in the equations coming from $L$, since $\Phi^+$and $\tilde \Phi^+$ are not independent degrees of freedom, we expect more equation on $\Phi^-$ and $\tilde \Phi^-$. Indeed, supersymmetry constraints the derivative of, for instance, $K^{mnp1}$, which does not appear in (\ref{Kfirst})-(\ref{dkt12l}). Using the extra equations that we present in Appendix \ref{App:extraK}, which involve the combination of $\Phi^-$ and $\tilde{\bar \Phi}^-$ with an opposite sign as that of (\ref{DPhi}), we get the following set of equations
\beq \begin{aligned}
\label{dDPhi-decoupled}
e^{-2A} d_H (e^{2A} \Phi^-) &=d(A+\phi) \wedge s(\bar{\Tilde \Phi}^-) - \Lambda_0 F \ , \\
e^{-2A} d_{-H} (e^{2A} \tilde \Phi^-) &=d(A+\phi) \wedge s(\bar{ \Phi}^-) - \bar \Lambda_0' F\ .
\end{aligned}
\eeq
where $d_{-H}=d+H\wedge$ and $s$ was defined in (\ref{Mukai}).
\newline{}
Note that the R-R fluxes only enter the equations through $\Lambda$, which is zero in the standard ${\cal N}=2$ ansatz. Their contribution also goes away in the equation for the even (odd) spinors if $\eta^1$ and $\tilde \eta^2$ are parallel (orthogonal). 
\newline{}
The additional equations on $\Lambda^-$ coming from (\ref{DL}) are the following 
\beq
\begin{aligned}
d \im \Lambda_1&=0 \ , \qquad &d \im \Lambda_5=0 \ , \\
e^A d(e^{-A} \re \Lambda_1)& = i \DPhi_2 \llcorner F_4 \ , \quad
&d \re \Lambda_5= \langle F, \DPhi^+ \rangle \ ,\\
\nabla_{(m|} (e^{-A} \re \Lambda_{|n)})&=0 \ , &
\end{aligned}
\eeq
plus the algebraic constraint
\beq
\re \Lambda_1 \llcorner H =0
\eeq
while from (\ref{DK}) we get additionally on $\Lambda^+$
\beq \begin{aligned}
d(e^{A-\phi} \Lambda_0)&=0 \ , \\
e^{-(A+\phi)} d(e^{A+\phi} \DLambda^+_{(+)4})&=-F\wedge \DPhi^-|_5 \ , \\
e^{A+\phi} *d(e^{-(A+\phi)} \DLambda^+_{(-)4})&=-i(*F)\llcorner \DPhi^-|_1 \ ,
\end{aligned}
\eeq
where we have defined
\beq
\DLambda^+_{(\pm)}=\Lambda^+\pm \bar \Lambda'^+
\eeq
Let us make a few comments before we go on to the discussion. First, note that the equations do not look exactly like a pair of ${\cal N}=1$ equations of the form (\ref{koerber1})-(\ref{koerber3}). This is because the ${\cal N}=2$ EGG formulation selects the pure spinors $\Phi$ and $\tilde \Phi$, instead of $\Lambda^1$, $\Lambda^2$, defined in (\ref{otherspinors}), which would be the natural ones from the ${\cal N}=1$ point of view. In other words, the present equations are the natural ones when one thinks of ${\cal N}=2$ backgrounds in terms of an $SU(6)$ structure on the exceptional generalized tangent space, and not in terms of a pair of $SU(7)$ structures.  Then, we notice that the equations for the pure spinors involve the $H$-twisted differential, while the R-R fluxes appear on the RHS only when $\Lambda_0, \Lambda_1$ are not-zero, or in other words when (at least one of) the spinors $\tilde \eta^I$ is not zero. In the case $\tilde \eta^I=0$, we get that $\Phi^+$ and $e^{2A} \Phi^-$ are $d_H$ closed, that $A=-\phi$ and the R-R fluxes should obey certain algebraic constraints. One  solution within this class is the generalized K\"ahler solution \cite{Hitchin} (previously called ``bihermitian geometry" \cite{GHR}), where $F=A=\phi=0$, and the two pure spinors are $H$-twisted closed.       

\section{Discussion} \label{discussion}
We have found the conditions on the twisted derivative of the structures $L$ and $K_a$ required by compactifications to four-dimensional Minkowski vacua preserving ${\cal N}=2$ supersymmetry. As expected from doing variations on the triplet of Killing prepotentials in (\ref{PaE7}), ${\cal N}=2$ supersymmetry requires these structures to be twisted closed. Two subtleties arise, though. The first one is that there is one component of ${\cal D} L$ and one component of ${\cal D} K$ which are not zero. Massaging these two equations as much as possible, we were able to write the obstruction to twisted closure in terms respectively of a single R-R and NS-NS flux contracted with an appropriate $L$ and $K$. These combinations are not set to zero by the other equations. The fact that these components of the twisted derivatives of $L$ and $K_a$ are not zero is surprising, but does not contradict with the expectation coming from four-dimensional supergravity, since they involve derivatives of components of $L$ and $K_a$ that need to be projected out in order to obtain a standard ${\cN}=2$ off-shell formulation (see \cite{GLSW} for more details). The second subtlety is that there are explicit powers of the dilaton appearing in certain equations, though we could make sense of them considering how the dilaton appears when embedding $GL(6, \mathbb R)$ into $SL(8, \mathbb R)$. Furthermore, these powers of the dilaton appear uniformly in the GCG counterpart equations.    
\newline{}
In the language of gauged supergravity, ${\cN}=2$ conditions arise from requiring that the matrices $S$, $W$ and $N$, appearing respectively in the susy variations of the gravitini, gaugini and hyperini vanish. The conditions obtained in \cite{HLV}-\cite{Louis} from setting to zero $S$ and $W$, should be equivalent to our conditions on ${\cal D}K_a$, while the ones coming from setting $N=0$, should  translate into our conditions on ${\cal D}L$. It would be nice to have an explicit check of this, as was done in \cite{BC},\cite{KM} in the case of ${\cN}=1$ vacua and the equations on the pure spinors of generalized complex geometry.  
\newline{}
By parameterizing the $SU(8)$ spinors in terms of $SU(4)$ ones, we decomposed $L$ and $K_+$ into $O(6,6)$ pure spinors. The structure $L$ contains the difference between two even pure spinors $\Phi$ and $\tilde \Phi$, while  $K_+$ contains their odd counterparts, and they each have additional degrees of freedom. The ${\cal N}=2$ 
equations for $\Phi$ and $\tilde \Phi$ involve the $H$-twisted differential $d-H \wedge$, while the R-R fluxes appear on the right hand side, multiplying the extra degrees of freedom. These equations simplify considerably in the ``standard ${\cal N}=2$ ansatz", where $L$ and $K_+$ contain just $\Phi^+$ and $\Phi^-$ respectively. In this case the R-R fluxes completely decouple (and should obey some algebraic constraints), while the pure spinors are twisted integrable, \textit{i.e.} they describe a generalized K\"ahler structure.

\subsection*{Acknowledgements}
We would like to thank Hagen Triendl and Daniel Waldram for extremely valuable discussions. 
This work is supported by the DSM CEA-Saclay and by the ERC Starting Independent Researcher Grant 259133 -- ObservableString. FO would like to acknowledge Consorzio Ferrara Ricerche (CFR) for financial support.

\appendix

\section{${\cal N}=1$ vacua in Generalized Geometry} \label{app:N=1}
In this Appendix we show the differential conditions on the algebraic structures (the pure spinors in GCG, $L$ and $K_a$ in EGG) required by ${\cal N}=1$ vacua on warped Minkowski space (\ref{warped}) in the presence of 
NS-NS and R-R fluxes.
The preserved spinor can be parameterized
within the $\cN=2$ spinor ansatz (\ref{fulla}) by a doublet $n_I=( a, \bar b)$
such that the supersymmetry preserved is given by $\epsilon=n_I \epsilon^I$. One can then always make a redefinition\footnote{\label{foot:tildeeta} This redefinition is $\eta^1+\tilde \eta^2 \to \eta^1$, $\tilde \eta^1+\eta^2 \to \eta^2$).}  such that the preserved spinor is 
\beq \label{thetaN=1}
\epsilon=\xi_- \otimes (\theta^1 + \theta^2) + c.c. \ , \qquad {\rm with} \ 
\theta^1= \left(\begin{array}{c} a \eta^1_+ \\ 0 \end{array}\right)       \ , \    \theta^2= \left(\begin{array}{c} 0 \\ \bar b \eta^2_{-} \end{array}\right)
\eeq 
and we take $|\eta^1|^2=|\eta^2|^2=1$ (while $|a|$ and $|b|$ are related to the warp factor, as we will see, and we have that for Minkowski vacua $|a|=|b|$).
The vector $n_I$ distinguishes a $U(1)_R \subset SU(2)_R$ such that any triplet can be written in terms of a 
$U(1)$ complex doublet and a $U(1)$ singlet by means of the vectors 
\begin{align} \label{zr}
(z^+,z^-,z^3)&=n_I (\sigma^a)^{IJ} n_J=(a^2,-\bar b^2, -2a\bar b) \ , \\
(r^+,r^-,r^3)&=n_I  (\sigma^a)^I{}_J \bar n^J=(ab,\bar a \bar b, |a|^2-|b|^2) \nn \ .
\end{align}
Using these vectors, one can extract respectively an $\cN=1$ superpotential ${\cal W}=z^a {\cal P}_a$ and an $\cN=1$ D-term ${\cal D}=r^a  {\cal P}_a$ from the triplet
of Killing prepotentials ${\cal P}_a$ that give the potential in the $\cN=2$ theory. This triplet of prepotentials is nicely written in terms of the Mukai pairing in $O(6,6)$ and the symplectic invariant in $\Es7$ between the two algebraic structures. The conditions for ${\cal N}=1$ vacua can be obtained  from extremizing the superpotential and setting the D-term  to zero. We will now give the GCG description, and then go on to its exceptional counterpart. 

\subsection{${\cal N}=1$ vacua in GCG}
Using the pure spinors of GCG introduced in (\ref{purespinors}), the triplet of Killing prepotentials reads in type IIA \cite{GLW}
\beq \label{Pa}
{\cal P}_+=\langle \Phi^+, d_H \Phi^- \rangle \ , \quad {\cal P}_-=\langle \Phi^+, d_H \bar \Phi^- \rangle \ , \quad
{\cal P}_3=-\langle \Phi^+, F^+ \rangle \ .
 \eeq
The conditions for Minkowski vacua preserving $\cN=1$ supersymmetry in the presence or NS-NS and R-R fluxes have been
obtained in \cite{GMPT2} in the language of GCG, using the ten-dimensional gravitino and dilatino variations (written respectively in (\ref{svariations}), (\ref{svariations2})), and in \cite{KM,BC} were shown to arise from the four-dimensional effective action as well. For the case $|a|=|b|$, which is the case in Minkowski compactifications with orientifold planes, they read
\begin{align} 
d_H(e^{2A}\Phi'^{+})&=0\label{koerber1}\\
d_H(e^{A}\mbox{Re}\Phi'^{-})&=0 \label{koerber2}\\
d_H( e^{3A} \mbox{Im}\Phi'^-)&=*e^{4A} s(F^+) \label{koerber3}
\end{align}
where 
\beq \label{phiprime}
\Phi'^+=2 a\bar b \, \Phi^+ \ , \quad \Phi'^-=2 a b \, \Phi^- \ .
\eeq
Finally, $\cN=1$ supersymmetry requires
\beq \label{norma}
|a|^2+|b|^2=e^{A} \ .
\eeq
 Conditions (\ref{koerber1})-(\ref{koerber3}) can be understood as coming from F and D-term equations. Equation (\ref{koerber2}) corresponds to imposing ${\cal D}=0$, while (\ref{koerber1}) and (\ref{koerber3})  come respectively from variations of the superpotential with respect to $\Phi^-$ and $\Phi^+$. \\
The susy condition in (\ref{koerber1}) says that the GACS ${\cal J}^+$ (see \eqref{Jgen}) is twisted integrable, and furthermore that the canonical bundle is trivial, and therefore the required manifold is a twisted Generalized Calabi-Yau. The other GACS featured in (\ref{koerber2})-(\ref{koerber3})  is ``half integrable", \textit{i.e.} its real part is, while the non-integrability of the imaginary part 
is due to the R-R fluxes.

\subsection{${\cal N}=1$ vacua in EGG}
The expression for the triplet of Killing prepotentials in terms of $L$ and $K_a$, the relevant algebraic structures in EGG, is given in \eqref{PaE7}. The complex and real vectors $z^a$, $r^a$ defined in (\ref{zr}) are used to build a complex and a real combination of the triplet $K_a$, that we will $K'_1$ and $K'_+$, which are the ones that will enter respectively in the superpotential and D-term. More precisely, we define 
\begin{align}
L'&\equiv e^{2A} L \ , \qquad  \nn \\
\ K'_1&\equiv  e^A r^a K_a=e^A K_1 \ ,  \\
K'_+&\equiv  e^{3A} z^a K_a=e^{3A} (K_3+iK_2) \ . \nn 
\end{align}
\newline{}
In the language of EGG, $\mathcal{N}=1$ supersymmetry requires requires for $L$, 
\beq
\cd L' \big|_{\bf{133}}=0 \label{N=1EGG1} \  ,
\eeq
for ${\cal D}{K}'_1|_{\bf 56}$
\begin{align}
(\cd K'_{1})'^{mn}=0,  \qquad \qquad {(\cd K'_{1})}_{mn}&=0  \ , \nn \\
\quad( \cd K'_{1})'^{12} =0, \qquad  \qquad {(\cd K'_{1})}_{12}&=0 \ , \label{N=1EGG2}\\
(\cd K'_{1})'^{m2}=0, \qquad \qquad  {(\cd K'_{1})}_{m1}&=0 \ \nn , 
\end{align}
and for ${\cal D}{ K}'_+|_{\bf 56}$
\begin{align}
(\cd K'_{+})'_{mn} - i {(\cd K'_{+})}_{mn}&=0  \ , \nn \\
\quad (\cd K'_{+})'_{12} - i {(\cd K'_{+})}_{12}&=0 \ , \label{N=1EGG3}\\
(\cd K'_{+})'^{m2}&=0\ . \nn 
\end{align}
The remaining  components of ${\cal D} K$ (all with one internal index) are proportional to derivatives of the dilaton and warp factor as follows 
\begin{align} 
&( \cd K'_{1})'^{m1} =4e^{-2A} \partial_pA K'_+{}^{mp}, \quad \  {(\cd K'_{1})}_{m2}=-4 e^{-2A} \partial_p A \, (2K'_{+}{}^{p}{}_{m12}+i \delta^p{}_m K'_+{}^1{}_2) \label{vectorDK1}  ,\\
&(\cd (e^{-\phi} K'_{+}))'^{m1} = -4i e^{-\phi} g^{mp} \partial_p A  K'_+{}^1{}_2 \ , \  {(\cd (e^{2A-\phi} K'_{+}))}_{m2}= - e^{2A-\phi} H_{mpq}K'_+{}^{12pq}\, ,  \label{vectorDK} \\
& {(\cd (e^{-4A+\phi} K'_{+}))}_{m1}= 0  \ . 
\end{align}
The equations for $L$, ${K}_3'$ and ${K_+'}$ in (\ref{N=1EGG1})-(\ref{N=1EGG3}) are respectively the EGG version of (\ref{koerber1}), (\ref{koerber2}) and (\ref{koerber3}).
The vectorial equations are a combination of (\ref{koerber1})-(\ref{koerber3}) plus (\ref{norma}).

\newpage{}
\section{$SL(8,{\mathbb R})$ components of $L$ and $K_a$ in terms of pure spinors} \label{SL8LK}
To obtain the $SL(8,\mathbb R)$ components of $L$ and $K_a$, we use (\ref{pw1}) and (\ref{pw2}). Then we want to split the $SL(8,\mathbb R)$ index into a $GL(6,\mathbb R)$ and an $SL(2,\mathbb R)$ index. For that, we use the embedding of $GL(6,\mathbb R)$ into $SL(8,\mathbb R)$
given by the following metric (for more details see \cite{GO1})
\beq \label{8dmetric}
\hat g_{ab}=\left(\begin{array}{ccc} g^{-1/4} g_{mn} & 0 & 0 \\ 0 & g^{-1/4} e^{-2\phi} & 0 \\ 0 & 0 & g^{3/4} e^{2\phi} \end{array} \right) 
\eeq
as well as  
the following decomposition for Cliff(8) gamma matrices in terms of the Cliff(6) ones $\gamma^m$ 
\begin{align} \label{gammabasis}
\Gamma^m&= g^{1/8} \ \ \ \ \,  \gamma^m \otimes \sigma_3 \nonumber \\
\Gamma^1&= g^{1/8} e^{\phi} \, \, \ \ 1_6\otimes \sigma_1 \\
\Gamma^2&= g^{-3/8} e^{-\phi} \,  1_6\otimes \sigma_2 \nonumber \ .
\end{align}
This gives, for the $12$ components of $L$ for example
\begin{align}
L^{12} &= -i \frac{\sqrt{2}}{2} (\Phi^+_0 - \bar{\tilde \Phi}_0^+) \ , \nn \\
L_{12}&= - i \frac{\sqrt{2}}{2} g^{1/2}  (\Phi^+_0 - \bar{\tilde \Phi}_0^+) \nn \ 
\end{align}
where the subscript $0$ denotes the zero-form piece of the polyform corresponding to the $O(6,6)$ spinor through the Clifford map (\ref{Cliffordmap}), and we have used the fact that $L$ transforms in the ${\bf 56}$ representation weighted by a power  of $g^{1/4}\simeq(\Lambda^6 T^*M)^{1/2}$.  We now note that given the factor of $g^{1/2}$ in $L_{12}$, this transforms as a six-form, namely the Hodge star of the zero-form. Using additionally that the pure spinors are imaginary anti self-dual, \textit{i.e.} $* \Phi^{\pm}=-i \Phi^{\pm}$, we can write
\beq
L_{12}= -\frac{\sqrt{2}}{2}(\Phi^+_6 + \bar{\tilde \Phi}^+_6) \nn \ .
\eeq 
We proceed similarly for the other components of $L$ and get 
\beq \label{compLSL8}
\begin{array}{ll}
L^{12} = -i \frac{\sqrt{2}}{2} (\Phi^+_0 - \bar{\tilde \Phi}_0^+) \ , &L_{12}= -\frac{\sqrt{2}}{2}(\Phi^+_6 + \bar{\tilde \Phi}^+_6) \ ,  \\
L^{mn}=-i \frac{\sqrt{2}}{2} \hat{\epsilon}^{mnpqrs} (\Phi^+_4 + \bar{\tilde \Phi}^+_4)_{pqrs} 		\ , 	\quad &L_{mn}=  \frac{\sqrt{2}}{2} (\Phi^+_2 - \bar{\tilde \Phi}^+_2)_{mn}  \\
L^{m1}=i e^{\phi} \sqrt{2} \hat{\epsilon}^{mnpqrs} ({\rm Re} \Lambda^-_5)_{npqrs}    \ , &L_{m1}=- i e^{-\phi} \sqrt{2}  ({\rm Im} \Lambda^-_1)_{m}     \\
L^{m2}=	- i e^{-\phi} \sqrt{2}  ({\rm Re} \Lambda^-_1)^{m}    \ , &L_{m2}= -i e^{\phi} \sqrt{2} g^{1/2}  ({\rm Re} \Lambda^-_1)_{m} 
\end{array}
\eeq
where $\hat \epsilon$ is a numeric totally antisymmetric tensor (\textit{i.e.} with values $\pm1, 0$), such that $L^{mn}$, for example, transforms as a 4-form.
\newline{}
For $K_+$, weighting by a factor $g^{1/2}$, we get the following components 
\beq \label{compKSL8}
\begin{array}{ll}
K_+{}^2{}_1=\frac{i}{4}  e^{-2\phi} (\Lambda_0^+-\bar{\Lambda'}_0^+) \ , & \\
K_+{}^1{}_2=-\frac{1}{4} g^{1/2} e^{2\phi} (\Lambda_6^++\bar{\Lambda'}_6^+) \ , & K_+^{mn12}=\frac{i}{8} \hat \epsilon^{mnpqrs}(\Lambda_4^++\bar{\Lambda'}_4^+)_{pqrs} \ , \\
K_+{}^m{}_n=-\frac14 g^{1/2} (\Lambda_2^++\bar{\Lambda'}_2^+)^m{}_n  \ , & K_+^{mnp2}=\frac{i}8 e^{-\phi} \hat \epsilon^{mnpqrs}(\Phi_3^--\bar{\tilde \Phi}_3^-)_{qrs} \ , \\
K_+{}^m{}_1=\frac{i}{4} e^{-\phi} \hat \epsilon^{mnpqrs}(\Phi^-_5 +\bar{\tilde{\Phi}}_5^-)_{npqrs} \ , \, & K_{+\, mnp1}=\frac{1}{8} e^{-\phi} (\Phi_3^--\bar{\tilde \Phi}_3^-)_{mnp} \\
K_+{}^2{}_m=-\frac{i}{4} e^{-\phi}(\Phi^-_1 +\bar{\tilde{\Phi}}_1^-)_{m} \ , & \, K_{+}^{mnp1}=\frac{i}{8} e^{\phi} g^{1/2} \hat \epsilon^{mnpqrs} (\Phi_3^-+\bar{\tilde \Phi}_3^-)_{qrs} \ , \\
K_+{}^1{}_m=\frac{1}{4} e^{\phi} g^{1/2} (\Phi^-_1 -\bar{\tilde{\Phi}}_1^-)_{m} \, & \\
K_+{}^m{}_2=\frac{1}{4} e^{\phi} g^{1/2} \hat \epsilon^{mnpqrs} (\Phi^-_5 -\bar{\tilde{\Phi}}_5^-)_{npqrs} \, & 
\end{array}
\eeq
where we have multiplied the whole ${\bf 133}$ representation by a factor $g^{1/2}$. To obtain the components of $K_3$, we first write it in terms of pure spinors as
\begin{equation}
K_{3}= \left(\begin{array}{cc}
\Phi^{1+}- \tilde \Phi^{2+} & \Lambda^{1-} - \Lambda^{2-T}\\
\bar{\Lambda}^{1-T} - \bar{\Lambda}^{2-} & \bar{\tilde \Phi}^{1+} - \bar{\Phi}^{2+} 
\end{array}\right) \nn
\eeq
where  we have defined
\begin{align} \label{otherspinors}
\Phi^{1+} & = e^{-\phi} \eta^1_+ \eta^{1 \dagger}_+ \ , \qquad \Phi^{2+}  =e^{-\phi}  \eta^2_+ \eta^{2 \dagger}_+ \ , \qquad \Lambda^{1-}  =e^{-\phi} \eta^1_+ \tilde \eta^{1 \dagger}_- \ , \\
\tilde \Phi^{1+} & =e^{-\phi} \tilde \eta^1_+\tilde \eta^{1 \dagger}_+  \ , \qquad \tilde \Phi^{2+}  =e^{-\phi} \tilde \eta^2_+\tilde \eta^{2 \dagger}_+\ ,  \qquad   \Lambda^{2-}  =e^{-\phi} \eta^2_+ \tilde \eta^{2 \dagger}_-  \ . \nn \ 
\end{align}
Then, for the $SL(8,\mathbb{R})$ components of $K_3$ we just need to make the following replacements in (\ref{compKSL8}) 
\beq
\begin{array}{cc}
\Lambda^+ \to \Phi_1^+-\tilde \Phi_2^+ \ , \quad &\, \, \Phi^- \to \Lambda_1^-+s(\Lambda_2^-) \ , \\
\bar \Lambda'^+ \to \bar{\tilde{\Phi}}_1^+-\bar{\Phi}_2^+ \ , \quad &\bar{\tilde{\Phi}}^- \to -s(\bar{\Lambda}_1^-)-\bar \Lambda_2^-
\end{array}
\end{equation}
where the operation $s$ on forms, which corresponds to minus (plus) the transposed of the bispinors, was defined in (\ref{Mukai}).

\section{$SL(8,\mathbb{R})\subset E_{7(7)}$ tensor product representations}\label{E7}
The $\SLE$ decomposition of the tensor products is the following.\\
The symplectic invariant $\bf{56}\times\bf{56}\big|_{\bf{1}}$ reads
\beq \label{symplSLE}
{\cal S}(\nu,\tilde \nu)= \nu'^{ab} \tilde \nu_{ab}-\nu_{ab} \tilde \nu'^{ab}
\eeq
The $\bf{56}\times\bf{56}\big|_{\bf{133}}$ reads 
\begin{align} \label{56x56=133SLE}
(\nu\cdot \tilde \nu)^{a}_{\,\,\,b}&=(\nu'^{ca}\tilde \nu_{cb}-\frac{1}{8}\delta^{a}_{\,\,\,b}\nu'^{cd}\tilde \nu_{cd})+(\tilde \nu'^{ca}\nu_{cb}-\frac{1}{8}\delta^{a}_{\,\,\,b}\tilde \nu'^{cd}\nu_{cd})\\
(\nu\cdot \tilde \nu)_{abcd}&=-3(\nu_{[ab}\tilde \nu_{cd]} - \frac{1}{4!} \epsilon_{abcdefgh} \nu'^{ef} \tilde \nu'^{gh}) \nn 
\end{align}
The $\bf{56}\times\bf{133}\big|_{\bf{56}}$ is
\begin{align} \label{56x133=56SLE}
(\nu \cdot \mu)^{ab}&=\mu^{a}_{\,\,\,c}\nu'^{cb}+\mu^{b}_{\,\,\,c}\nu'^{ac}+\star\mu^{abcd}\nu_{cd}\\
(\nu \cdot \mu)_{ab}&=-\mu^{c}_{\,\,\,a}\nu_{cb}-\mu^{c}_{\,\,\,b}\nu_{ac}-\mu_{abcd}\nu'^{cd} \nn
\end{align}
where $\star \mu$ is the 8-dimensional Hodge dual, while the adjoint action on the adjoint $\bf{133}\times\bf{133}\big|_{\bf{133}}$ gives
\begin{align} \label{133x133=133SLE}
(\mu \cdot \mu')^{a}_{\,\,\,b}&=(\mu^{a}_{\,\,\,c}\mu'^{c}_{\,\,\,b}-\mu'^{a}_{\,\,\,c}\mu^{c}_{\,\,\,b})+\frac{1}{3}(\star \mu^{acde}\mu'_{bcde}-\star\mu'^{acde}\mu_{bcde})\\
(\mu \cdot \mu')_{abcd}&=4(\mu^{e}_{\,\,\,[a}\mu'_{bcd]e}+\mu'^{e}_{\,\,\,[a}\mu_{bcd]e}) \nn 
\end{align}
The $\bf{56} \times \bf{133}\big|_{\bf{912}}$ is 
\begin{align} \label{56x133=912SLE}
(\nu \cdot \mu)^{ab}&=(\nu'^{ac}\mu^{b}_{\,\,\,c}+\nu'^{bc}\mu^{a}_{\,\,\,c}) \nn \\
(\nu \cdot \mu)_{ab}&= (\nu_{ac}\mu^{c}_{\,\,\,b}+\nu_{bc}\mu^{c}_{\,\,\,a}) \nn \\
(\nu \cdot \mu)^{abc}{}_{d}&=-3(\nu'^{[ab}\mu^{c]}_{\,\,\,\,b}-\frac{1}{3}\nu'^{e[a}\mu^{b}_{\,\,\,e}\delta^{c]}_{\,\,\,d})+2(\nu_{ed} \star\mu^{abce}+\frac{1}{2}\nu_{ef}\star\mu^{ef[ab}\delta^{c]}_{\,\,\,d})\\
(\nu \cdot \mu)_{abc}{}^d&=-3(\nu_{[ab}\mu^{d}_{\,\,\,c]}-\frac{1}{3}\nu_{e[a}\mu^{e}_{\,\,\,b}\delta^{d}_{\,\,\,c]})-2(\nu'^{ed} \mu_{abce}+\frac{1}{2}\nu'^{ef}\mu_{ef[ab}\delta^{d}{}_{c]}) \nn 
\end{align}
The $\bf{912}\times\bf{56}\big|_{\bf{133}}$ gives
\begin{align} \label{912x56=133SLE}
(\phi \cdot \nu)^{a}_{\,\,\,b}&=(\nu'^{ca}\phi_{cb}+\nu_{cb}\phi'^{ca})-(\nu_{cd}\phi'^{cda}{}_{b}-\nu'^{cd}\phi_{cdb}{}^a)\\
(\phi \cdot \nu)_{abcd}&=-4( \phi_{[abc}{}^e\nu_{d]e}-\frac{1}{4!}\epsilon_{abcdm_1m_2m_3m_4}\phi'^{m_1m_2m_3}{}_{e}\nu'^{m_4 e}) \nn 
\end{align}
and finally $\bf{912}\times\bf{133}\big|_{\bf{56}}$ is
\begin{align} \label{912x133=56SLE}
(\phi \cdot \mu)^{ab}&=-(\phi'^{ac}\mu^{b}_{\,\,\,c}-\phi'^{bc} \mu^{a}_{\,\,\,c})-2\phi'^{abc}{}_{d}\mu^{d}{}_{c}\nonumber\\&+\frac{2}{3}(\phi_{m_1m_2m_3}{}^a \star\mu^{m_1m_2m_3 b}-\phi_{m_1m_2m_3}{}^b \star\mu^{m_1m_2m_3 a})\\
(\phi \cdot \mu)_{ab}&=(\phi_{ac}\mu^{c}{}_{b}-\phi_{bc}\mu^{c}_{\,\,\,a})-2 \phi_{abc}{}^d \mu^{c}{}_{d}\nonumber\\&-\frac{2}{3}(\phi'^{m_1m_2m_3}{}_{b}\, \mu_{m_1m_2m_3a}-\phi'^{m_1m_2m_3}{}_{a} \, \mu_{m_1m_2m_3b}) \label{signrev}
\end{align}

\section{Supersymmetric variations for the $\mathcal{N}=2$ spinor anstaz}\label{susy}
The supersymmetry transformations of the fermionic fields of type IIA, namely the gravitino $\psi$ and dilatino $\lambda$ read, in the democratic formulation \cite{demo} 
\begin{equation} 
\delta \psi_M = \nabla_M \epsilon +\frac{1}{4} \sla{H_M} {\cal P} \epsilon + \frac{1}{16} e^{\phi}  
\sum_n  \sla \! {F^{(10)}_{n}} \, \Gamma_{M} {\mathcal P}_n \, \epsilon  \, , \label{svariations}
\end{equation} 
\begin{equation}  
\delta \lambda = \left(\sla{\partial} \phi + \frac{1}{2} \sla \! H {\mathcal P}\right) \epsilon 
+ \frac{1}{8} e^{\phi}  
\sum_n  (5-n)  \sla \! {F^{(10)}_{n}} \,  
{\cal P}_n  \epsilon  \, .\label{svariations2}
\end{equation} 
where $\psi$, $\lambda$ and $\epsilon$ are a doublet of spinors of opposite chirality, as in (\ref{fulla}), and ${\cal P}=-\sigma^3$,  ${\cal P}_n= (-\sigma_3)^{n/2} \sigma_1$ act on the doublet.
\newline{}
We use the standard decomposition of ten-dimensional gamma matrices
\beq
\gamma^{(10)}_{\mu}= \gamma_{\mu} \otimes 1 \, , \ \ \gamma^{(10)}_m=\gamma_5
\otimes \gamma_m \, ,
\eeq
and the Poincare invariant ansatz for the R-R fluxes
\beq 
\label{eq:rrfs} 
F^{(10)}_{2n} = F_{2n} + {\rm vol}_4 \wedge {\tilde F}_{2n-4} \, \quad {\rm where}\  {\tilde F}_{2n-4} = 
(-1)^{Int[n]}   *_6 {F}_{10-2n} \ .
\eeq 
 Using (\ref{gammabasis}), we notice that  ${\cal P}=i \Gamma^{12}$, ${\cal P}_0={\cal P}_4=\Gamma^1$, ${\cal P}_2={\cal P}_6=-i \Gamma^2 $, $\gamma^m {\cal P}_0=-i \Gamma^{2m}$ and  
 $\gamma^m  {\cal P}_2=\Gamma^{1m}$ and obtain from the internal components of the gravitino variation  that $\mathcal{N}=2$ supersymmetry requires for the internal spinors in the spinor ansatz (\ref{suvar}) 
\begin{equation} \label{intgravitino}
\delta \psi_m=0 \  \Leftrightarrow \  \nabla_m \theta^I=-\frac{i}{8}H_{mnp}\Gamma^{np12}\theta^I+\frac{e^{\phi}}{8}\sla{\FI}\Gamma_m\theta^I \ , 
\end{equation}
while from the external gravitino variation, we get
\begin{equation} \label{extgravitino}
\delta \psi_\mu=0 \  \Leftrightarrow \ i \sla{\PE} A \, \theta^I +\frac{e^{\phi}}{4}\sla{\FE}\theta^{I}=0 \ , \quad 
\end{equation}
and from the dilatino variation
\begin{equation} \label{dilatino}
\delta \lambda=0 \  \Leftrightarrow   i \sla{\PE} \phi \, \theta^I +\frac{1}{12}H_{mnp}\Gamma^{mnp}\theta^I+\frac{e^{\phi}}{4}\sla{\FD}\theta^I=0 \ .
\end{equation}
In these equations we have defined
\begin{equation} \label{FI}
\sla{F_i}=-i {\sla}{F_h} \Gamma^2+{\sla}{F_a} \Gamma^1 \ , \quad \sla{\FE}= \sla{F_h} \Gamma^1-i {\sla}{F_{a}} \Gamma^{2} \, ,  \quad 
\sla{\FD}=(5-n) \sla{\FE}  \ . 
\eeq
in terms of the ``hermitean" and ``antihermitean" pieces of $F$, namely
\beq
F_h = \frac12(F+s(F))=F_{0}+F_{4} \ , \quad F_a =\frac12(F-s(F))= F_{2}+F_{6}
\eeq
and a slash means
\beq
{\sla}{F}_{(n)}=\frac{1}{n!} F_{i_1...i_n} \Gamma^{i_1...i_n} \ . 
\eeq
Finally
\beq
\sla{\PE} A =\partial_m A \, \Gamma^{m12} \ .
\eeq

\section{${\cal D} L$ and ${\cal D} K$ versus $\mathcal{N}=2$ supersymmetry} \label{app:DLDKvssusy}

\subsection{$\cD L$}\label{n2L}
Multiplying Eqs. (\ref{intgravitino}), (\ref{extgravitino}) and (\ref{dilatino}) (coming respectively from the internal and external gravitino and dilatino)  on the right by $\epsilon_{IJ} e^{-\phi} \theta^J$, we get the following equations on $L$
\begin{align}
(\Delta_m L)^{\alpha\beta}&\equiv\nabla_mL^{\alpha\beta}+\partial_m\phi L^{\alpha\beta}+\frac{i}{4}H_{mnp}(\Gamma^{np12}L)^{\alpha\beta}-\frac{e^{\phi}}{4}(\sla{F_i}\Gamma_m L)^{\alpha\beta}=0\,,\label{igln2}\\
(\Delta_e L)^{\alpha\beta}&\equiv i\partial_p A( \Gamma^{p12}L)^{\alpha\beta}+\frac{e^{\phi}}{4}(\sla{F_e}L)^{\alpha\beta}=0\,,\label{egln2}\\
(\Delta_d L)^{\alpha\beta}&\equiv i\partial_p \phi (\Gamma^{p12}L)^{\alpha\beta}+\frac{1}{12}H_{pqr}(\Gamma^{pqr}L)^{\alpha\beta}+\frac{e^{\phi}}{4}(\sla{F_d}L)^{\alpha\beta}=0\,,\label{dln2}
\end{align}
\newline{}
We can also multiply (\ref{extgravitino}) and (\ref{dilatino}) on the left by $\epsilon_{IJ} e^{-\phi} \theta^J$, and get
\begin{align}
(L\Delta_e )^{\alpha\beta}&\equiv i\partial_p A(L \Gamma^{p12})^{\alpha\beta}-\frac{e^{\phi}}{4}(L\sla{F_e})^{\alpha\beta}=0\,,\\
(L\Delta_d )^{\alpha\beta}&\equiv i\partial_p \phi (L\Gamma^{p12})^{\alpha\beta}-\frac{1}{12}H_{pqr}(L\Gamma^{pqr})^{\alpha\beta}-\frac{e^{\phi}}{4}(L\sla{F_d})^{\alpha\beta}=0\, . 
\end{align}
We will also need the ``transposed" of the equation coming from internal gravitino, namely 
\begin{align}
 (L\Delta_m)^{\alpha\beta}&\equiv \nabla_mL^{\alpha\beta}+\partial_m\phi L^{\alpha\beta}-\frac{i}{4}H_{mnp} (L\Gamma^{np12})^{\alpha\beta}+\frac{e^{\phi}}{4}(L\sla{F_i}\Gamma_m)^{\alpha\beta}=0\,.
 \end{align}
 \newline{}
 Given $L$ and product of gamma matrices $\Gamma^{a_1\dots a_i}$ we will make use of the following type of combinations 
\beq \label{commut}
 \Tr \left( [\Gamma^{a_1\dots a_i} ,\Deltaphi]L\right) = \Tr \left((\Gamma^{a_1\dots a_i} \Deltaphi -  {\Deltaphi}\Gamma^{a_1\dots a_i} ) L \right)=\Tr \left(\Gamma^{a_1\dots a_i} \Deltaphi L- L {\Deltaphi}\Gamma^{a_1\dots a_i} \right) \ .
\eeq
and similarly for the anticommutator. 
 \newline{}
Multiplying these equations by appropriate combinations of gamma matrices, we recover the combinations involved in (\ref{Lfirst})-(\ref{Llast}).  Unless otherwise specified, we will take 
\be
l_d=1\, 
\ee
in the very last step of the following equations.
\newline{}
We start from (\ref{lt1m})
\begin{align}
0=&\frac{\sqrt{2}}{4} \Tr \left[ \Gamma^{12} \Delta_m  L+i   \Gamma_{m} l_d \Delta_dL \right] \nonumber\\
=&\nabla_m L^{12}+\partial_m\phi L^{12}-l_d\partial_m\phi L^{12}  \nn \\
&+ \frac{i}{4}H_{mnp}L^{np}(-1+l_d)\nn\\
&+\frac{e^{\phi}}{4}\left[iF_0 (1-5l_d)+(*F_6)(1-l_d)\right]L^1_{\,\,\,m}  \nn\\
&+\frac{e^{\phi}}{4}\left[F_{mp}(-1-3l_d)+i(*F_4)_{mp}(1-l_d)\right]L^{2p}\nn\\
=&\nabla_{m}L^{12}-ie^{\phi}F_0 L^{1}{}_{m}-e^{\phi}F_{mp}L^{2p}\nn\\
=&\,(\cD L)^{1}{}_{m}\,.
\end{align}
Similarly, we have
\begin{align}
  0=& \frac{\sqrt{2}}{4} \Tr \left[ - \Gamma^{mp}\Delta_p L +i\Gamma^{m12} l_d \Delta_dL \right] \nonumber\\
=& -\nabla_p L^{mp}+(l_d-1)\partial_p\phi L^{mp}+\nn\\
&  +\frac{i}{4}(3-l_d)(*H)^{mpq}L_{pq}\nn\\
&+\frac{e^{\phi}}{4}\left(iF_{0}(5-5l_d)-(*F_6)(5-l_d)\right)L^{m}{}_{2}\nn\\
& +\frac{e^{\phi}}{4}\left(F^{mp}(3-3l_d)-i(*F_4)^{mp}(3-l_e-l_d)\right)L^{1}{}_{p}\nn\\
& -\nabla_p L^{mp}+\frac{i}{2}(*H)^{mpq}L_{pq}-e^{\phi}(*F_6)L^{m}{}_2-e^{\phi}(*F_4)^{m}{}_{p}L^{1p}\,,\nn\\
=&\,(\cD L)^{m}{}_2\, 
\end{align}
and
\begin{align}
 0&= \frac{\sqrt{2}}{4} \Tr \left[ \frac{3i}{2}\Gamma_{[mn|} \Delta_{|p]} L +\frac{1}{2}  \Gamma_{mnp}{}^{12} l_d \Delta_dL  \right]\nonumber\\
=& +\frac{3}{2}i\nabla_{[m|} L_{|np]}+\frac{3}{2}i (1-l_d) \partial_{[m|}\phi L_{|np]}\nn\\
&+\frac{1}{4}(3-l_d)H_{mnp}L^{12}+\frac{3}{4}(1-l_d)(*H)_{q[mn|}L^{q}{}_{|p]}\nn\\
&-\frac{3}{8}e^{\phi}\left(iF_{[mn|}(1+3l_d)+(*F_4)_{[mn|}(1-l_d)\right)L^{1}{}_{|p]}\nn\\
&-\frac{e^{\phi}}{8}\left(i(*F_2)_{mnpq}(3-3l_d)+F_{mnpq}(3+l_d)\right)L^{2q}\nn\\
=&\frac{3}{2}i\nabla_{[m|} L_{|np]}+\frac{1}{2}H_{mnp}L^{12} -\frac{3}{2}ie^{\phi}F_{[mn|}L^{1}{}_{|p]}-\frac{e^{\phi}}{2}F_{mnpq}L^{2q}\nn\\
=&\,(\mathcal{D}L)_{mnp2}\,.
\end{align}
\newline{}
Consider now
\begin{align}
 0&= \frac{\sqrt{2}}{4} \Tr   \left[- \Gamma^{1m} \Delta_m L+i   \Gamma^{2} l_d\Delta_dL \right]  \nonumber\\
=&-\nabla_m (L^{1m}) -(1-l_d) \partial_m \phi e^{\phi} L^{1m}\nn\\
&-\frac{e^{\phi}}{4}\left(iF_0(-6+5l_d)+(*F_6)(6-l_d)\right)L^{12}\nn\\
&-\frac{e^{\phi}}{8}\left(F_{mn}(2-3l_d)+i(*F_4)_{mn}(-2-l_d)\right)L^{mn}
\end{align}
Choosing this time $l_d=2$ we recover
\begin{align}
0=&-\nabla_m (L^{1m})+\partial_m \phi e^{\phi} L^{1m} -e^{\phi}(iF_0+(*F_6))L^{12}+e^{\phi}(F_{mn}+i(*F_4)_{mn})L^{mn}\nn\\
&=e^{\phi} \left(\mathcal{D}(e^{-\phi}L)\right)^{1}{}_{2}
\label{DL12}
\end{align}
We are then left with two equations. 
Using only the internal gravitino constraint we get
\begin{align}
 0=& \frac{\sqrt{2}}{4}i \Tr \left[  \Gamma_{1[n|}\Delta_{|m]}  L\right] \nn\\
 =&-i\nabla_{[m|} L_{|n]1}-i\partial_{[m}\phi L_{n]1}+\frac{1}{2}H_{mnp}L^{2p}\nn\\
=&e^{-\phi} (\mathcal{D}(e^{\phi}L))_{mn12}\label{DLmn12}
\end{align}
For $(\cD L)^{n}{}_{m}$ we have on one hand
\begin{align}
0=& \frac{\sqrt{2}}{4}\Tr\left[\Delta_m L\Gamma^{n2} \right]=\nabla_{m}L^{n2}+\partial_{m}\phi L^{n2}-\frac{i}{2}H^{n}{}_{mp}L^{1p}\nn\\
&+\frac{e^{\phi}}{4}\left[iF_0-(*F_6)\right]L^{n}{}_{m}\nn\\
&+\frac{e^{\phi}}{4}\left[F^{n}{}_{m}-i(*F_4)^{n}{}_{m}\right]L^{12}\nn\\
&-\frac{e^{\phi}}{8}\left[(*F_2)^{n}{}_{mpq}-iF^{n}{}_{mpq}\right]L^{pq}
\end{align}
and on the other hand we can use
\begin{align}
0=& \frac{\sqrt{2}}{4}\Tr\left[L\Delta_m \Gamma^{n2}\right]=\nabla_{m}L^{n2}+\partial_{m}\phi L^{n2}+\frac{i}{2}H^{n}{}_{mp}L^{1p}\nn\\
&+\frac{e^{\phi}}{4}\left[iF_0+(*F_6)\right]L^{n}{}_{m}\nn\\
&+\frac{e^{\phi}}{4}\left[-F^{n}{}_{m}-i(*F_4)^{n}{}_{m}\right]L^{12}\nn\\
&-\frac{e^{\phi}}{8}\left[-(*F_2)^{n}{}_{mpq}-iF^{n}{}_{mpq}\right]L^{pq}
\end{align}
By comparing the two, one recovers the following constraint
\beq
\frac{i}{2}H^{n}{}_{mp}L^{1p}+\frac{e^{\phi}}{4}\left[(*F_6)L^{n}{}_{m}-F^{n}{}_{m}L^{12}+\frac{1}{2}(*F_2)^{n}{}_{mpq}L^{pq}\right]=0
\eeq
Consider then the following combination using the commutator defined in (\ref{commut})
\begin{align}
 0=&\frac{\sqrt{2}}{4}\Tr \left[\Delta_m \Gamma^{n2}L+i[\Delta_d l_d+\Delta_e l_e, \Gamma_{m}{}^{n1}]L\right]\nn\\
=&\nabla_{m}L^{n2}+\partial_{m}((1+l_d) \phi+l_eA)L^{n2}\nn\\
&+i\frac{e^{\phi}}{4}\Big[F_0(1+5l_d+l_e)L^{n}{}_{m} -(*F_4)^{n}{}_{m}(1+l_d+l_e)L^{12}\nn\\
&+\frac{1}{2}F^{n}{}_{mpq}(1-l_d-l_e)L^{pq}\Big]\nn \ .
\end{align}
The following choice for $l_d$ and $l_e$ makes the equation look as simple a possible
\beq
l_d=0\,,\quad l_e=-1\,
\eeq
for which we obtain
\begin{align}
0=&\nabla_{m}L^{n2}+\partial_{m}(\phi-A)L^{n2}+i\frac{e^{\phi}}{4}F^{n}{}_{mpq}L^{pq}\nn\\
 =&e^{-(\phi-A)} (\cD(e^{(\phi-A)}L))^{n}{}_{m}+i\frac{e^{\phi}}{4}F^{n}{}_{mpq}L^{pq}\,.
\end{align}

\subsection{${\cal D} K$}
\label{n2K}
We need the hermitean conjugate of Eq. \eqref{intgravitino}, namely
\begin{equation}
\nabla_m\bar \theta_{J}=\frac{i}{8}H_{mnp} \bar \theta_J \Gamma^{np12} -\frac{e^{\phi}}{8} \bar \theta_J \Gamma_m \sla{\FI} \ .\label{thetaadj}
\end{equation}
Multiplying \eqref{intgravitino} by $\sigma_a \bar \theta_J$, and (\ref{thetaadj}) by $\theta^I \sigma_a$, we get the following condition (for any $a$)
\begin{align}
\Deltam \Kp\equiv &\nabla_m  \Kp{}^{\alpha}{}_{\beta} -\partial_m(A-\phi)\Kp{}^{\alpha}{}_{\beta} +\frac{i}{8}H_{mnp}[\Gamma^{np12}\Kp-\Kp\Gamma^{np12}]^{\alpha}{}_{\beta} \nn \\
& -\frac{e^{\phi}}{8}[\sla{F_i}\Gamma_m \Kp-\Kp\Gamma_m\sla{F_i}]^{\alpha}{}_{\beta}=0 \label{intgravitinoK}\, .
\end{align}
Using a similar trick on the external gravitino and dilatino equations (\ref{extgravitino}) and (\ref{dilatino}), we also get 
\begin{align}
(\Deltamu \Kp)^{\alpha}{}_{\beta}&\equiv i\partial_m A [\Gamma^{m12}K_1]^{\alpha}{}_{\beta}+\frac{e^{\phi}}{4}[\sla{F_e}\Kp]^{\alpha}{}_{\beta}=0\label{egra1} \ , \\
(\Deltaphi \Kp)^{\alpha}{}_{\beta}&\equiv i\partial_m \phi [\Gamma^{m12}\Kp]^{\alpha}{}_{\beta}+\frac{1}{12}H_{mpq}[\Gamma^{mpq}\Kp]^{\alpha}{}_{\beta}+\frac{e^{\phi}}{4}[\sla{F_d}K]^{\alpha}{}_{\beta}=0 \label{dila1}\ .
\end{align}
and their ``transposed" versions
\begin{align}
(\Kp \Deltamu )^{\alpha}{}_{\beta}&\equiv i\partial_m A [\Kp\Gamma^{m12}]^{\alpha}{}_{\beta}-\frac{e^{\phi}}{4}[\Kp\sla{F_e}]^{\alpha}{}_{\beta}=0\label{egra2} \ , \\
(\Kp {\Deltaphi} )^{\alpha}{}_{\beta}&\equiv i\partial_m \phi [\Kp\Gamma^{m12}]^{\alpha}{}_{\beta}-\frac{1}{12}H_{mpq}[\Kp\Gamma^{mpq}]^{\alpha}{}_{\beta}-\frac{e^{\phi}}{4}[\Kp\sla{F_d}]^{\alpha}{}_{\beta}=0 \label{dila2}\,.
\end{align}
\newline{}
We sketch in the following how conditions (\ref{DK}) arise from supersymmetry. We first look at the $mn$ components. We have
\begin{align}
0=&-\frac{i}{4} \Tr[-\Gamma^{mnp2}\Delta_p \Kp+ i\Gamma^{mn1} (n_e \Delta_e   + n_d \Delta_d)\Kp] & \nn \\
=& -2\nabla_p \Kp^{mnp2}+2\partial_p(A-\phi)\Kp^{mnp2}-2 i \partial_r(n_e A +n_d \phi) g^{r[m}\Kp^{2|n]} +2\partial_p(n_d\phi+n_e A)\Kp^{mnp2}\nn \\
& +\frac{1}{2}(1-n_d) H^{mn}{}_p \Kp^{1p} - in_d H_{pq}{}^{[m|} \Kp^{1pq|n]} +\frac{1}{2} (3-n_d) (*H)^{mnp} \Kp^2{}_p \nn \\
&+ \frac{e^{\phi}}{4} \left[F_0(-4+n_e+5n_d)- i (*F_6) (n_e-n_d)\right] \Kp^{mn}\nn \\
&+\frac{e^{\phi}}{4}\left[i F^{mn}(n_e+3n_d)-(*F_4)^{mn}(4+n_e+n_d)\right]\Kp^{12}\nn\\
&+\frac{e^{\phi}}{8}\left[ i (*F_2)^{mn}{}_{pq}(n_e+3n_d)-F^{mn}{}_{pq}(n_e+n_d)\right]\Kp^{pq}\nn\\
&+e^{\phi}\left[F^{[m|}{}_{p}(-2+n_e+3n_d)-i(*F_4)^{[m|}{}_p(n_e+n_d)\right]\Kp^{p12|n]}\label{n2dkmnu}
\end{align}
and
\begin{align}
 0=& -\frac{1}{4}\Tr[-2 \Gamma^{2}{}_{[n|} \Delta_{|m]}\Kp  +i \Gamma_{mn1} ( n_e  \Delta_e + n_d \Delta_d)\Kp )] \nn\\
=&-2\nabla_{[m|}\Kp^{2}{}_{|n]}+2 \partial_{[m|}(A -\phi)\Kp^{2}{}_{|n]}+2 \partial_{[m|}(n_e A +n_d \phi)\Kp^{2}{}_{|n]} -2i\partial_p(n_d\phi+n_e A)\Kp^{2mnp}\nn\\
&-i\frac{n_d}{2} H_{mnp}\Kp^{1p}-(1-n_d)H_{[m|pq}\Kp^{1pq}{}_{|n]}-i\frac{n_d}{2}(*H)_{mnp}\Kp^{2p}\nn\\
&+\frac{e^{\phi}}{4}\left(iF_0(n_e+5n_d)+(*F_6)(2+n_e-n_d)\right)\Kp_{mn}\nn\\
&+\frac{e^{\phi}}{4}\left(-F_{mn}(2+n_e+3n_d)-i(*F_4)_{mn}(n_e+n_d)\right)\Kp^{12}\nn\\
&+\frac{e^{\phi}}{8}\left(-(*F_2)_{mnpq}(-2+n_e+3n_d) -iF_{mnpq}(n_e+n_d)\right)\Kp^{pq}\nn\\
&+e^{\phi}\left(iF_{[m|p}(n_e+3n_d)+(*F_4)_{[m|p}(n_e+n_d)\right)\Kp^{q12}{}_{|n]}\label{n2dkmnd} \ .
\end{align}
Consider first \eqref{n2dkmnu}. By choosing $n_d=1,\, n_e=-1$, we recover :
\begin{align} \label{resint}
&0=-2\Kp^{mnp2}-2i\partial_{[m|}(\phi-A)\Kp^2{}_{|n]}-iH_{[m|pq}\Kp^{1pq}{}_{|n]}+(*H)_{mnp}\Kp^{2p}\nn\\
&+\frac{e^{\phi}}{2}\big[-i(*F_6)\Kp^{mn}+(iF_{mn}-2(*F_4)_{mn})\Kp^{12}+\frac{1}{2}(*F_2)_{mnpq}\Kp^{12p}{}_{n}\big]\,.
\end{align}
The above equality can be further decoupled in terms of its real and imaginary part, the first of which reads\footnote{Actually in our conventions $K^{abcd}, K^{a}{}_b$ are purely imaginary, so (\ref{realpart}) corresponds to the imaginary part of (\ref{resint}).}
\begin{align} \label{realpart}
0&=-2\nabla_{p}\Kp^{mnp2}+(*H)^{mn}{}_{p}\Kp^{2p}-e^{\phi}(*F_4)^{mn}\Kp^{12}\,\nn\\
&=(\mathcal{D}\Kp)^{mn}
\end{align}
while the second yields the following equation
\begin{align}
0&=2\partial_{[m|}(A-\phi)\Kp^{2}{}_{|n]}-H_{[m|pq}\Kp^{1pq}{}_{|n]}\nn\\
&+\frac{e^{\phi}}{2}\big[F_{mn}\Kp^{12}-(*F_6)\Kp_{mn}+\frac{1}{2}(*F_2)_{mnpq}\Kp^{pq}\big]\,.
\end{align}
In a very similar fashion, we consider \eqref{n2dkmnd} for the same values $n_d=+1,\,n_e=-1$, which gives
\begin{align}
0&=-2\nabla_{[m|}\Kp^2{}_{|n]}-2i\partial_p(\phi-A)\Kp^{2}{}_{mn}{}^{p}-\frac{i}{2}(H_{mnp}\Kp^{1p}+(*H)_{mnp}\Kp^{2p})\nn\\
&+e^{\phi}\big[iF_0\Kp_{mn}-F_{mn}\Kp^{12}+2iF_{[m|p}\Kp^{12p}{}_{|n]}\big]\,.
\end{align}
once more decoupling the real from the imaginary part we respectively recover 
\begin{align}
0&=-2\nabla_{[m|}\Kp^2{}_{|n]}-e^{\phi}F_{mn}\Kp^{12}=(\cD\Kp)'_{mn} \ ,
\end{align}
and
\begin{align}
0&=+2\partial_p(A-\phi)\Kp^{2}{}_{mn}{}^{p}-\frac{1}{2}(H_{mnp}\Kp^{1p}+(*H)_{mnp}\Kp^{2p})\nn\\
&+e^{\phi}\big[F_0\Kp_{mn}+2F_{[m|p}\Kp^{12p}{}_{|n]}\big]\ .
\end{align}
Now consider the 12 components. We start from
\begin{align} \label{dk12up}
0=&-\frac14 \Tr\left[-\Delta_p \Kp \Gamma^{p1}-i\Gamma^2(n_d\Delta_d+n_e\Delta_e)\Kp\right]\nn\\
&-\nabla_p\Kp^{p1}+\partial_p(A-\phi)\Kp^{p1}-\partial_p(n_e A +n_d \phi)\Kp^{1p}-\frac{1}{2}(1-\frac{n_d}{3})H_{mnp}\Kp^{2mnp}\nn\\
&+\frac{e^{\phi}}{4}\left(iF_0(5n_d+n_e)+(*F_6)(6+n_e-n_d)\right) \Kp^{12} \nn\\
&-\frac{e^{\phi}}{4}\left(F_{mn}(-2+3n_d+n_e)+i\left(*F_4)_{mn}(n_e+n_d\right)\right) \Kp^{mn} 
\end{align}
which specialized once more for $n_d=1,\, n_e=-1$ gives
\begin{align}
0=&-\nabla_{p}\Kp^{p1}-\frac{1}{3}H_{mnp}\Kp^{2mnp}+e^{\phi}\big[+iF_0+(*F_6)\big]\Kp^{12}
\end{align}
which has a very intuitive decomposition in imaginary and real contributions:
\begin{align}
0=&-\nabla_{p}\Kp^{p1}-\frac{1}{3}H_{mnp}\Kp^{2mnp}+e^{\phi}(*F_6)\Kp^{12}=(\cD \Kp)'^{12}\,,\\
0=&e^{\phi}F_0\Kp^{12}=(\cD \Kp)^{12}\,.ÿ
\end{align}
We discuss in the following the remaining components.
\begin{align}
0=&-\frac{1}{4}\Tr\left[\Delta_m \Kp^{12}-i\{ \Delta_e n_e, \Gamma_m \} \Kp\right] \nn \\
=&-\nabla_m \Kp^{2}{}_{1}+\partial_m (A-\phi)\Kp^{2}{}_1 +n_e\partial_p A \Kp^{12}\nn\\
&+\frac{e^{\phi}}{4}(1-n_e)\left[ F_{mp}\Kp^{2p}+(*F_4)_{pq}\Kp^{2pq}{}_{m}-(*F_6)\Kp_{m1}\right]
\end{align}
which by taking $n_e=1$  simplifies to
\beq
0=-\nabla_{m}\Kp^{2}{}_1-\partial_m\phi \Kp^{2}{}_{1}=e^{-\phi}(\cD(e^{\phi}\Kp))_{m1}\,.
\eeq
For the components $(\cD \Kp)'^{m1}$ and $(\cD K)_{m2}$ we need to separate the R-R contributions from the rest.
For the first one, notice that
\begin{align}
0&=-\Tr\left[ [n_d\Delta_d+n_e\Delta_e,\Gamma_m]\Kp\right]=e^{\phi}\left(F_0 \Kp_{m1}-(*F_4)_{mp}\Kp^{2p}-F_{pq}\Kp^{2pq}{}_{m}\right)-8\partial_p A \Kp_{m}{}^{p12}\nn\\
&=\cF_{R-R} \big|'^{m1}-8\partial_p A \Kp^{mp12}\,.\label{rrm1}
\end{align}
We thus have
\begin{align}
0&=\frac{i}{4}\Tr \left[\Gamma^{mp12}\Delta_p \Kp +i[2\Delta_d-5\Delta_e,\Gamma^m]\Kp\right]\nn\\
&=-2\partial_p(-5A+2\phi)\Kp^{mp12}-2\partial_p(A-\phi)\Kp^{mp12}+2\nabla_p \Kp^{mp12}\nn \\
&=2\nabla_p \Kp^{mp12}-2\partial_p(-4A+\phi)\Kp^{mp12}\nn\\
&=2\nabla_p \Kp^{mp12}-(\cF_{R-R}\big|'^{m1}-8\partial_p A \Kp^{mp12})-2\partial_p(-4A+\phi)\Kp^{mp12}\nn\\
&=(\cD \Kp)'^{m1}-2\partial_{p}\phi \Kp^{mp12}=e^{\phi} (\mathcal{D}(e^{-\phi}\Kp))'^{m1}\,.
\end{align}
where on the third line we made explicit use of \eqref{rrm1}.\\
Then for $({\cD K})_{m2}$ the argument is similar. We first find the R-R piece in the connection in the following combination
\begin{align}
0&=-\Tr\left[ i \Gamma_{m}{}^{21}\Delta_e \Kp+\Delta_m \Kp\right] \nn \\
&=-4\partial_p A \Kp_{m}{}^{p}+(*F_6)\Kp_{m2}+F_{mp}\Kp^{1p}+(*F_4)_{pq}\Kp^{1pq}{}_{m}\nn\\
&=-4\partial_p A \Kp_{m}{}^{p}+\cF_{R-R}\big|_{m2}\, .
\end{align}
Consider then the following combination using the commutator introduced in (\ref{commut})
\begin{align}
0&=\frac{1}{4}\Tr\left[\Gamma^{p}{}_{m}\Delta_p \Kp+i[3\Delta_e-2\Delta_d, \Gamma_{m}{}^{12}\Kp]\right]\nn\\
&=-\nabla_p \Kp^{p}{}_{m}-\partial_p(3A-2\phi)\Kp^{p}{}_{m}+\partial_p(A-\phi)\Kp^{p}{}_{m} \nn \\
&=-\nabla_p \Kp^{p}{}_{m}-\partial_p (2A-\phi)\Kp^{p}{}_{m}\nn\\
&=-\nabla_p \Kp^{p}{}_{m}-H_{mpq}\Kp^{12pq}+H_{mpq}\Kp^{12pq}+(-4\partial_p A \Kp_{m}{}^{p}+\cF_{R-R}\big|_{m2})-\partial_p(2A-\phi)\Kp^{p}{}_{m}\nn\\
&=(\cD \Kp)_{m2}+\partial_p(2A+\phi)\Kp^{p}{}_{m}+H_{mpq}\Kp^{12pq}\nn\\
&=e^{2A+\phi} (\cD(e^{-(2A+\phi)}\Kp))_{m2}+H_{mpq}\Kp^{12pq}\,.
\end{align}

\subsubsection{Extra equations on $K$ required by susy}
\label{App:extraK}
In this section, we will use equations on the object
\beq \label{choice}
\Kpp=e^{3A} K \ .
\eeq
We will make use of (\ref{intgravitinoK}), coming from gravitino, which on $\Kpp$ has an additional factor of $3$ in front of the derivative of the warp factor. The following combinations are required to vanish by supersymmetry
\begin{align}
0=&\frac{i}{4}\Tr[\nabla_p\Kpp\Gamma^{mnp1}+i\Gamma^{mn2}(\nabla_d n_d+\nabla_e n_e)\Kpp]\nn\\
=&-2\nabla_p\Kpp^{mnp1}+2\partial_{p}(\phi-3A)\Kpp^{mnp1}-2i\partial^{[m|}(n_eA+n_d\phi)\Kpp^{1|n]}-2\partial_{p}(n_eA+n_d\phi)\Kpp^{1mnp}\nn\\
&-\frac{1}{2}(1+n_d)H^{mnp}\Kpp^{2}{}{p}+\frac{1}{2}(3+n_d)(*H)^{mnp}\Kpp^{1}{}_{p}\nn\\
&+\frac{e^{\phi}}{4}[iF_0(5n_d+n_e)-(*F_6)(-4+n_e-n_d)]\Kpp^{mn12}\nn\\
&-\frac{e^{\phi}}{8}[(*F_2)^{mn}{}_{pq}(3n_d+n_e)+iF^{mn}{}_{pq}(n_e+n_d)]\Kpp^{pq12}\nn\\
&+\frac{e^{\phi}}{2}[iF^{p[m|}(3n_d+n_e)-(*F_4)^{p[m|}(-2+n_e+n_d)]\Kpp^{|n]}{}_{p}
\end{align}
Choosing $n_d=-1, n_e=3$ we get from the real part 
\begin{align}
0=&-2\nabla_p\Kpp^{mnp1}+2\partial_{p}(-3A+\phi)\Kpp^{mnp1}+2\partial_p(3A-\phi)\Kpp^{mnp1}+(*H)^{mnp}\Kpp^{1}{}_{p}\,\nn\\
=&-2\nabla_p\Kpp^{mnp1}+(*H)^{mnp}\Kpp^{1}{}_{p}\,.
\end{align}
Analogously
\begin{align}
0=&-\frac{1}{4}\Tr[-2\nabla_{[m|}\Kpp\Gamma^{1}{}_{|n]}\Kpp+i\Gamma_{mn}{}^{2}(\nabla_d n_d+\nabla_e n_e)\Kpp]\nn\\
=&2\nabla_{[m|}\Kpp^{1}{}_{|n]}+2\partial_{[m|}(\phi-3A)\Kpp^{1}{}_{|n]}\nn\\
&-2\partial_{[m|}(n_eA+n_d\phi)\Kpp^{1}{}_{|n]}+2i\partial_{p}(n_eA+n_d\phi)\Kpp^{1mnp}\nn\\
&+i\frac{n_d}{2}[H_{mnp}\Kpp^{2p}-(*H)_{mnp}\Kpp^{2p}]+(1+n_d)H_{[m|pq}\Kpp^{2pq}{}_{|n]}\nn\\
&+\frac{e^{\phi}}{2}[F_0(2+5n_d+n_e)+i(*F_6)(n_e-n_d)]\Kpp_{mn12}\nn\\
&+\frac{e^{\phi}}{4}[i(*F_2)_{mnpq}(3n_d+n_e)-F_{mnpq}(-2+n_e+n_d)]\Kpp^{pq12}\nn\\
&+\frac{e^{\phi}}{4}[F_{[m|p}(3n_d+n_e)+i(*F_4)_{[m|p}(n_e+n_d)]\Kpp^{p}{}_{|n]}\,.
\end{align}
which again from the real part and for $n_d=-1, n_e=3$, leads to
\be
0=-2\nabla_{[m|}\Kpp^{1}{}_{|n]}\,.
\ee
Finally
\begin{align}
0=&-\frac{1}{4}\Tr[-\Gamma^{m}{}_{2}\nabla_m\Kpp-i\Gamma^{1}(n_d\nabla_d+n_e\nabla_e)\Kpp]\nn\\
=&-\nabla_{m}\Kpp^{m}{}_{2}+\nabla_{m}(\phi-3A)\Kpp^{m}{}_2+\partial_{m}(n_d\phi+n_eA)\Kpp^{m}{}_2\nn\\
&+\frac{1}{2}(1+\frac{n_d}{3})H_{pqr}\Kpp^{1pqr}-ie^{\phi}(3n_d+n_e)F_{pq}\Kpp^{pq12}\nn\\
&-\frac{e^{\phi}}{12}[2-n_e-n_d]F_{mnpq}\Kpp^{mnpq}\,.
\end{align}
Using again $n_d=-1,n_e=3$ we get that the real component of the above equation is
\be
0=-\nabla_m \Kpp^{m}{}_{2}+\frac{1}{3}H_{mnp}\Kpp^{1mnp} \ .
\ee 
These three equations give the ``complement" of (\ref{dDPhiminus}) that allows us to decouple (though not completely) the equations for $\Phi^-$ and those for $\tilde \Phi^-$.

\end{document}